\documentclass[10pt,apj,tighten,iop]{emulateapj}

\usepackage{amsmath}   
\usepackage{graphicx}
\usepackage{tabulary}
\usepackage{epsfig}
\usepackage{amssymb}
\usepackage{multirow}
\usepackage{appendix}
\usepackage{natbib}
\usepackage{lineno}
\usepackage{wasysym}
\usepackage{color}         
\usepackage[pdftitle={Detecting Extrasolar Moons akin to Solar System Satellites with an Orbital Sampling Effect}, colorlinks = true, breaklinks = true, citecolor = blue, linkcolor = blue, urlcolor = blue, pdfauthor = {Ren\'{e} Heller}]{hyperref}
\def\wu#1{\sqrt{{#1}\!\;\,}^{\!\;\!\rule[-0.13ex]{.0453em}{.6ex}}\;\!}

\shorttitle{Detecting Extrasolar Moons akin to Solar System Satellites with an Orbital Sampling Effect}

\shortauthors{Ren\'e Heller}

\submitted{Received 2014 January 10; accepted 2014 March 23; published 2014 April 29 in The Astrophysical Journal}

\begin{document}

\title{DETECTING EXTRASOLAR MOONS AKIN TO SOLAR SYSTEM SATELLITES \\ WITH AN ORBITAL SAMPLING EFFECT}

\author{Ren\'e Heller\altaffilmark{1,2}}
\affil{Origins Institute, McMaster University, Hamilton, ON L8S 4M1, Canada \\\href{mailto:rheller@physics.mcmaster.ca}{rheller@physics.mcmaster.ca}}

\altaffiltext{1}{Department of Physics and Astronomy, McMaster University}
\altaffiltext{2}{Postdoctoral fellow of the Canadian Astrobiology Training Program}

\begin{abstract}
Despite years of high accuracy observations, none of the available theoretical techniques has yet allowed the confirmation of a moon beyond the solar system. Methods are currently limited to masses about an order of magnitude higher than the mass of any moon in the solar system. I here present a new method sensitive to exomoons similar to the known moons. Due to the projection of transiting exomoon orbits onto the celestial plane, satellites appear more often at larger separations from their planet. After about a dozen randomly sampled observations, a photometric orbital sampling effect (OSE) starts to appear in the phase-folded transit light curve, indicative of the moons' radii and planetary distances. Two additional outcomes of the OSE emerge in the planet's transit timing variations (TTV-OSE) and transit duration variations (TDV-OSE), both of which permit measurements of a moon's mass. The OSE is the first effect that permits characterization of multi-satellite systems. I derive and apply analytical OSE descriptions to simulated transit observations of the \textit{Kepler} space telescope assuming white noise only. Moons as small as Ganymede may be detectable in the available data, with M stars being their most promising hosts. Exomoons with the 10-fold mass of Ganymede and a similar composition (about 0.86 Earth radii in radius) can most likely be found in the available \textit{Kepler} data of K stars, including moons in the stellar habitable zone. A future survey with \textit{Kepler}-class photometry, such as \textit{Plato 2.0}, and a permanent monitoring of a single field of view over 5 years or more will very likely discover extrasolar moons via their OSEs.
\end{abstract}


\section{Context and Motivation}
\label{sec:context}

Although more than 1000 extrasolar planets have been found, no extrasolar moon has been confirmed. Various methods have been proposed to search for exomoons, such as analyses of the host planet's transit timing variation \citep[TTV;][]{1999A&AS..134..553S,2007A&A...470..727S}, its transit duration variation \citep[TDV;][]{2009MNRAS.392..181K,2009MNRAS.396.1797K}, direct photometric observations of exomoon transits \citep{2011ApJ...743...97T}, scatter analyses of averaged light curves \citep{2012MNRAS.419..164S}, a wobble of the planet-moon photocenter \citep{2007A&A...464.1133C}, mutual eclipses of the planet and its moon or moons \citep{2007A&A...464.1133C,2009PASJ...61L..29S,2012MNRAS.420.1630P}, excess emission of transiting giant exoplanets in the spectral region between 1 and 4\,$\mu$m \citep{2004AsBio...4..400W}, infrared emission by airless moons around terrestrial planets \citep{2009AsBio...9..269M,2011ApJ...741...51R}, the Rossiter-McLaughlin effect \citep{2010MNRAS.406.2038S,2012ApJ...758..111Z}, microlensing \citep{2002ApJ...580..490H}, pulsar timing variations \citep{2008ApJ...685L.153L}, direct imaging of extremely tidally heated exomoons \citep{2013ApJ...769...98P}, modulations of radio emission from giant planets \citep{2013arXiv1308.4184N}, and the generation of plasma tori around giant planets by volcanically active moons \citep{2014ApJ...785L..30B}. Recently, \citet{2012ApJ...750..115K} started the \textit{Hunt for Exomoons with Kepler} (HEK),\footnote{\href{http://www.cfa.harvard.edu/HEK}{www.cfa.harvard.edu/HEK}} the first survey targeting moons around extrasolar planets. Their analysis combines TTV and TDV measurements of transiting planets with searches for direct photometric transit signatures of exomoons.

Exomoon discoveries are supposed to grant fundamentally new insights into exoplanet formation. The satellite systems around Jupiter and Saturn, for example, show different architectures with Jupiter hosting four massive moons and Saturn hosting only one. Intriguingly, the total mass of these major satellites is about $10^{-4}$ times their planet's mass, which can be explained by their common formation in the circumplanetary gas and debris disk \citep{2006Natur.441..834C}, and by Jupiter opening up a gap in the heliocentric disk during its own formation \citep{2010ApJ...714.1052S}. The formation of Earth is inextricably linked with the formation of the Moon \citep{1976LPI.....7..120C}, and Uranus' natural satellites indicate a successive ``collisional tilting scenario'', thereby explaining the planet's unusual spin-orbit misalignment \citep{2012Icar..219..737M}. Further interest in the detection of extrasolar moons is triggered by their possibility to have environments benign for the formation and evolution of extrasolar life \citep{1987AdSpR...7..125R,1997Natur.385..234W,2013AsBio..13...18H}. After all, astronomers have found a great number of super-Jovian planets in the habitable zones (HZs) of Sun-like stars \citep{IJA:9150120}.

In this paper, I present a new theoretical method that allows the detection of extrasolar moons. It can be applied to discover and characterize multi-satellite systems and to measure the satellites' radii and orbital semi-major axes around their host planet, assuming roughly circular orbits. This assumption is justified because eccentric moon orbits typically circularize on a million year time scale due to tidal effects \citep{2011ApJ...736L..14P,2013AsBio..13...18H}. The method does not depend on a satellite's direction of orbital motion (retrograde or prograde), and it relies on high-accuracy averaged photometric transit light curves. I refer to the physical phenomenon that generates the observable effect as the Orbital Sampling Effect (OSE). It causes three different effects in the phase-folded light curve, namely, (1) the photometric OSE, (2) TTV-OSE, and (3) TDV-OSE. Similarly to the photometric OSE, the scatter peak method developed by \citet{2012MNRAS.419..164S}Ê\ makes use of orbit-averaged light curves. But I will not analyze the scatter. While the scatter peak method was described to be more promising for moons in wide orbits, the OSE works best for close-in moons. Also, with an orbital semi-major axis spanning $82\,\%$ of the planet's Hill sphere, the example satellite system studied by \citet{2012MNRAS.419..164S} would only be stable if it had a retrograde orbital motion \citep{2006MNRAS.373.1227D}.

\section{The Orbital Sampling Effect}
\label{sec:ose}

\begin{figure}[t]
  \centering
  \scalebox{0.18}{\includegraphics{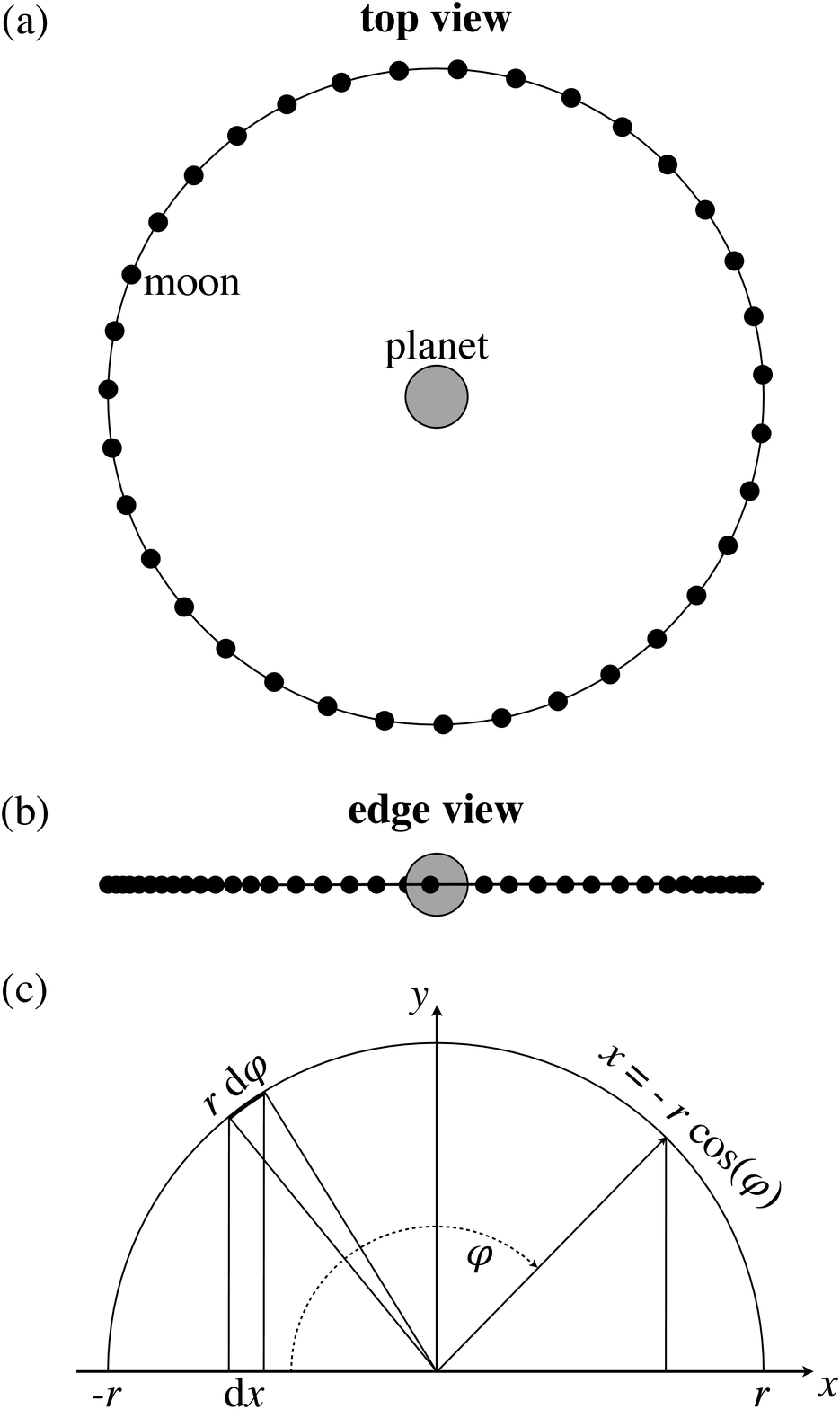}}\\
  \vspace{0.3cm}
  \caption{Geometry of a moon's OSE. Assuming a constant sampling frequency over one moon orbit (panel (a), an observer in the moon's orbital plane would recognize a non-uniform projected density distribution (panel (b)). All snapshots combined in one sequence frame, the moon is more likely to occur at larger separations $x$ from the planet. The probability distribution $P_\mathrm{s}(x)$ along the projected orbit can be constructed as $P_\mathrm{s}(x)=r \ \mathrm{d}\varphi/\mathrm{d}x$ (panel (c)).}
  \label{fig:geometry}
\end{figure}

\begin{figure}[t]
  \centering
  \scalebox{0.5}{\includegraphics{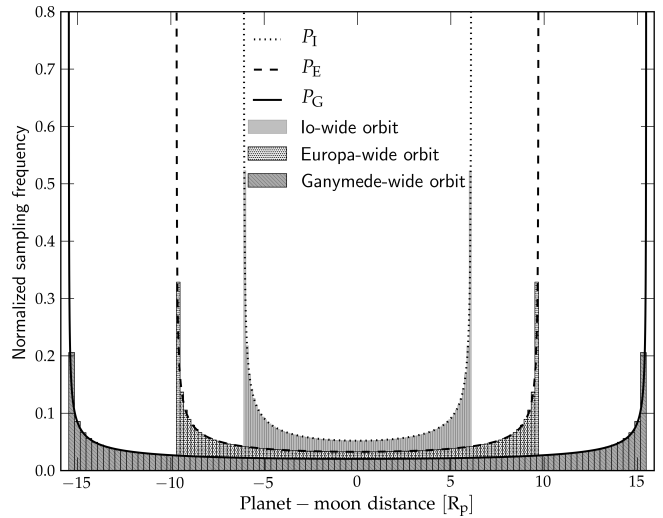}}
  \caption{Normalized sampling frequency (or probability density $P_\mathrm{s}(x)$) for three moons in an Io-wide, Europa-wide, and Ganymede-wide orbit in units of planetary radii. Bars show the results from a randomized numerical simulation, curves show the distribution according to Equation~(\ref{eq:ProbDens}). The integral, that is, the area under each curve equals 1. This explains why moons on tighter orbits have a higher sampling frequency at a given planet-moon distance.}
  \label{fig:ProbDens}
\end{figure}

\subsection{Probability of Apparent Planet-moon Separation}

Imagine a moon orbiting a planet on a circular orbit. In a satellite system with $n$ moons, this particular moon shall be satellite number $s$, and its orbital semi-major axis around the planet be $a_{\mathrm{ps}}$.\footnote{Speaking about the orbital geometry of a planet-moon binary as depicted in Figure~\ref{fig:geometry}, I will refer to the moon's orbital radius around the planet as $r$. In case of a multi-satellite system, I designate the planet-satellite orbital semi-major axis of satellite number $s$ as $a_\mathrm{ps}$.} Imagine further looking at the system from a top view position, such that the satellite's apparent path around the planet forms a circle. With a given, but arbitrary sampling frequency you take snapshots of the orbiting moon, and once the moon has completed one revolution, you stack up the frames to obtain a frame sequence. This sequence is depicted in panel (a) of Figure~\ref{fig:geometry}. An observer in the orbital plane of the satellite, taking snapshots with the same sampling frequency and stacking up a sequence frame from its edge view post, would see the moon's positions distributed along a line. The planet sits in the center of this line, which extends as far as the projected semi-major axis to either side (panel (b) of Figure~\ref{fig:geometry}). Due to the projection effect, the moon snapshots pile up toward the edges of the projected orbit. In other words, if the snapshots would be taken randomly from this edge-on perspective, then the moon would most likely be at an apparently wide separation from the planet. We can assume to observe most transiting exomoon systems in this edge view because the orbital plane of the planet-moon system should be roughly in the same plane as the orbital plane of the planet-moon barycenter around the star \citep{2011A&A...528A..27H}.

The likelihood of a satellite to appear at an apparent separation $x$ from the planet can be described by a probability density $P_\mathrm{s}(x)$, which is proportional to the ``amount'' of orbital path $r\times\mathrm{d}\varphi$, with $\varphi$ as the angular coordinate and $r$ as the orbital radius, divided by the projected part of this interval along the $x$-axis, $\mathrm{d}x$ (see panel (c) of Figure~\ref{fig:geometry}). With $x=-r\cos(\varphi)$, we thus have

\begin{align} \label{eq:OSE_basic} \nonumber
P_\mathrm{s}(x) &\propto \frac{r \ \mathrm{d}\varphi}{\mathrm{d}x} = r \frac{\mathrm{d}}{\mathrm{d}x} \arccos\left(\frac{-x}{r}\right) \\
        &= \frac{1}{{r^2}Ê\wu{\displaystyle 1 - \left(\frac{x}{r}\right)^2 } }  \ \ .
\end{align}

$P_\mathrm{s}(x)$ must fulfill the condition

\begin{equation} \label{eq:norm}
\int_{-r}^{+r} \mathrm{d}xÊ\ P_\mathrm{s}(x) = 1 \ \ ,
\end{equation}

\noindent
because it is a probability density, and the moon must be somewhere. With

\begin{equation}
\int_{-r}^{+r} \mathrm{d}xÊ\ \frac{1}{\wu{\displaystyle 1 - \left(\frac{x}{r}\right)^2 } } = \int_{-r}^{+r} \mathrm{d}xÊ\ r \ \frac{\mathrm{d}}{\mathrm{d}x} \arccos\left(\frac{-x}{r}\right) = {\pi}r
\end{equation}

\noindent
we thus have the normalized sampling frequency

\begin{equation}\label{eq:ProbDens}
P_\mathrm{s}(x) = \frac{1}{{\pi}r\wu{\displaystyle 1 - \left(\frac{x}{r}\right)^2 } } \ \ .
\end{equation}

In Figure~\ref{fig:ProbDens}, I plot Equation~(\ref{eq:ProbDens}) for a three-satellite system. The innermost moon with probability density $P_\mathrm{I}$ is in an orbit as wide as Io's orbit around Jupiter, that is, it has a semi-major axis of 6.1 planetary radii ($R_\mathrm{p}$). The central moon with normalized sampling frequency $P_\mathrm{E}$ follows a circular orbit $9.7\,R_\mathrm{p}$ from the planet, similar to Europa, and the outermost moon corresponds to Ganymede at $15.5\,R_\mathrm{p}$ and with probability density $P_\mathrm{G}$. The shaded areas illustrate a normalized suite of randomized measurements of the projected orbital separation $x$ that I have simulated with a computer. The lines follow Equation~(\ref{eq:ProbDens}), where $r$ is replaced by the respective orbital semi-major axis $a_{\mathrm{ps}}$, and they nicely match the randomized sampling. As the area under each curve equals $1$, the innermost moon has a higher probability to appear at a given position within the interval $[-a_\mathrm{pI},+a_\mathrm{pI}]$, with $a_\mathrm{pI}$ as the sky-projected orbital semi-major axis of the satellite in an Io-wide orbit, than any of the other moons. The normalized sampling frequency or probability density of apparent separation $P_\mathrm{s}(x)$ is independent of the satellite radius.

\subsection{The Photometric OSE}
\label{sub:photoOSE}

\subsubsection{The Photometric OSE in Averaged Transit Light Curves}
\label{subsec:OSEsignatures}

From the perspective of a data analyst, it is appealing that the OSE method does not require modeling of the orbital evolution of the moon or moons during the transit or between transits, that is, during the circumstellar orbit. Assuming that the satellite is not in an orbital resonance with the circumstellar orbital motion, the OSE will smear out over many light curves and always yield a probability distribution as per Equation~(\ref{eq:ProbDens}). What is more, this formula allows an analytic description of the actual effect in the light curve. In other words, once the phase-folded light curve is available after potential TTVs or TDVs -- induced by the moons or by other planets -- as well as red noise \citep{2013MNRAS.430.1473L} have been removed, the OSE can be measured with a simple fit to the binned data points. I refer to this effect as the photometric OSE.

\begin{figure}[t]
  \centering
  \scalebox{0.82}{\includegraphics[angle=90]{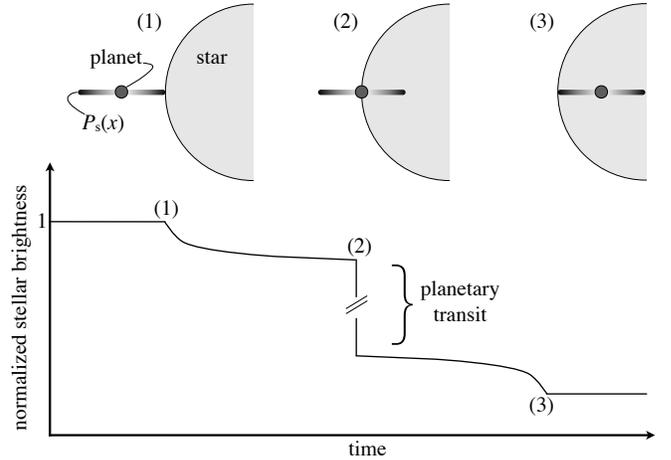}}\\
  \vspace{0.2cm}
  \caption{Photometric OSE during ingress. At epoch (1), a satellite's probability distribution $P_\mathrm{s}(x)$ along its circumplanetary orbit touches the stellar disk, from then on causing a steep (but small) decline in stellar brightness (see lower panel). As circumplanetary orbits with lower values of $P_\mathrm{s}(x)$ (visualized by lighter colors) enter the stellar disk, the brightness decrease weakens. At epochs (2), the planet enters the stellar disk and induces a dramatic decrease in stellar brightness, depicted by two slanted lines in the lower panel. From that point on, larger values of $P_\mathrm{s}(x)$ enter the stellar disk, so the slope of the stellar flux decrease as the OSE increases until the moon orbits have completely entered the stellar disk at epochs (3).}
  \label{fig:OSE}
\end{figure}

Yet, what does the effect actually look like? Figure~\ref{fig:OSE} visualizes the ingress of the planet-moon binary in front of the stellar disk from a statistical point of view. The satellite orbit is assumed to be coplanar with the circumstellar orbit, and the impact parameter (b) of the planet, corresponding to its minimal distance from the stellar center during the transit, in units of stellar radii, is zero. The thick shaded line drawn through the planet sketches the probability density $P_\mathrm{s}(x)$ and describes a smoothed-out version of the frame shown in panel (b) of Figure~\ref{fig:geometry}. If one were able to repeatedly take snapshots of a certain moment during subsequent transits (for example at epochs (1), (2), or (3) in this sketch), then the averaged positions of the moon would scatter according to this shaded distribution. I call the part right of the planet the right wing of $P_\mathrm{s}(x)$ (or $P^\mathrm{rw}$) and that part of $P_\mathrm{s}(x)$, which is left of the planet, the left wing of $P_\mathrm{s}(x)$ (or $P^\mathrm{lw}$).

As $P^\mathrm{rw}$ touches the stellar disk with its highest values, visualized by dark shadings in Figure~\ref{fig:OSE}, it causes a steep (though weak) decrease in the averaged transit light curve, as depicted in the bottom panel. The decreasing probability distribution then proceeds over the stellar disk and causes a declining decrease in stellar brightness. At epochs (2), the planet enters the stellar disk and triggers a major brightness decrease, commonly known from a planetary transit light curve. Moving on to point (3), that part of the probability function that actually moves into the stellar disk has increasingly higher values toward $P^\mathrm{lw}$, thereby inducing an increasing decline in stellar brightness. Note that the lower panel of Figure~\ref{fig:OSE} visualizes the averaged light curve. During each particular transit, the moon can be anywhere along its projected orbit around the planet, and the decrease in stellar brightness follows a completely different curve.

\subsubsection{Analytic Description of the Photometric OSE}
\label{subsec:OSEanalytic}

Equation~(\ref{eq:ProbDens}) allows for an analytic description of the photometric OSE. In Figure~\ref{fig:OSE}, the right wing of the probability distribution enters the stellar disk first, which corresponds to the right side of $P_\mathrm{s}(x)$ shown in Figure~\ref{fig:ProbDens}. Stellar light is blocked between $a_{\mathrm{ps}}$ to the right and $x'(t)$ to the left, where the latter variable describes the time dependence of the moving left edge of $P_\mathrm{s}(x)$. To calculate the amount of blocked stellar light due to the photometric OSE during the ingress of a one-satellite system ($F_\mathrm{OSE,in}^{(1)}$), I integrate $P_\mathrm{s}(x)$ from $x'(t)$ to $a_{\mathrm{ps}}$ and subtract this area from the normalized, apparent stellar brightness ($B_\mathrm{OSE}$), which equals 1 out of transit. The amplitude of the blocked light is given by $(R_\mathrm{s}/R_\star)^2$, where $R_\star$ is the stellar radius, and hence

\begin{align}\label{eq:OSE_in}\nonumber
F_\mathrm{OSE,in}^{(1)} \ =& \  \left(  \frac{R_\mathrm{s}}{R_\star} \right)^2 \int_{x'(t)}^{a_{\mathrm{ps}}} \mathrm{d}x \ P_\mathrm{s}(x) \\ \nonumber
                       =& \ \frac{1}{\pi} \left(  \frac{R_\mathrm{s}}{R_\star} \right)^2 \left[ \pi - \arccos\left(  \frac{-x'(t)}{a_{\mathrm{ps}}} \right)  \right] \\
                        & \hspace{2.13cm} _{\left\{\mathrm{for} \  -1 \ \leq \ \frac{-x'(t)}{a_{\mathrm{ps}}} \ \leq \ 1 \ \right\}}
\end{align}

\noindent
and $B_\mathrm{OSE,in}^{(1)}(t)\equiv1-F_\mathrm{OSE,in}^{(1)}(t)$ is the brightness during ingress. For $-x'(t)/a_{\mathrm{ps}}<-1$, that is, as long as the probability distribution of the moon has not yet touched the stellar disk, the $\arccos()$ term is not defined, so I set $F_\mathrm{OSE,in}^{(1)}~=~0$ in that case. To parameterize $x'(t)$, I choose a coordinate system whose origin is in the center of the stellar disk. Time is 0 at the center of the transit, so that $x'(t)=-R_\star-v_\mathrm{orb}t$, with $v_\mathrm{orb}=2{\pi}a_\mathrm{{\star}b}/P_\mathrm{{\star}b}$ as the circumstellar orbital velocity of the planet-moon barycentric mass $M_\mathrm{b}$, assumed to be equal to the circumstellar orbital velocity of both the planet and the moon, $P_\mathrm{{\star}b}= 2{\pi}\wu{a_\mathrm{{\star}b}^3/(G(M_\star+M_\mathrm{b}))}$ as the circumstellar orbital period of $M_\mathrm{b}$, $G$ as Newton's gravitational constant, and $a_\mathrm{{\star}b}$ as the orbital semi-major axis between the star and $M_\mathrm{b}$. To consistently parameterize the transit light curve and the OSE, the star-planet system must thus be well-characterized. Assuming that the planet is much more massive than the moon, taking $M_\mathrm{b}~{\approx}~M_\mathrm{p}$ is justified.

Once the whole probability density of the moon has entered the stellar disk, $-x'(t)/a_{\mathrm{ps}}>+1$ and the $\arccos()$ term is again not defined, so I take $F_\mathrm{OSE,in}^{(1)}~=~(R_\mathrm{s}/R_\star)^2$ in that case. During egress of the probability function, the moon, on average, uncovers a fraction $F_\mathrm{OSE,eg}^{(1)}$ of the stellar disk. Its mathematical description is similar to Equation~(\ref{eq:OSE_in}), except for $x'(t)=+R_\star-v_\mathrm{orb}t$. Again, $F_\mathrm{OSE,eg}^{(1)}=0$ before the egress of the probability function and $1$ after it has left the disk, so that the normalized, apparent stellar brightness becomes $B_\mathrm{OSE}^{(1)}(t)~=~1~-F_\mathrm{p}(t)-~F_\mathrm{OSE,in}^{(1)}(t)~+~F_\mathrm{OSE,eg}^{(1)}(t)$, with $F_\mathrm{p}(t)$ as the stellar flux masked by the transiting planet (see Appendix~\ref{sec:ingress}).

While Equation~(\ref{eq:OSE_in}) is valid for one-satellite systems, it can be generalized to a system of $n$ satellites by subtracting the stellar flux that is blocked subsequently by the integrated density functions via

\begin{equation}\label{eq:OSE_multi}
B_\mathrm{OSE}^{(n)}(t) = \ 
\begin{cases}
\ 1 - F_\mathrm{p}(t) - {\displaystyle \sum_{\mathrm{s}=1}^n} F_\mathrm{OSE,in}^{(\mathrm{s})}(t) \\
      \hspace{1.62cm} + {\displaystyle \sum_{\mathrm{s}=1}^n} F_\mathrm{OSE,eg}^{(\mathrm{s})}(t) \\
\hspace{4.1cm} \mathrm{for} \ |x_\mathrm{p}(t)| > R_\star \\
\\
\ 1 - F_\mathrm{p}(t) - {\displaystyle \sum_{\mathrm{s}=1}^n} F_\mathrm{OSE,in}^{(\mathrm{s})}(t) \\
      \hspace{1.62cm} + {\displaystyle \sum_{\mathrm{s}=1}^n} F_\mathrm{OSE,eg}^{(\mathrm{s})}(t) + A_\mathrm{mask} \\
\hspace{4.1cm} \mathrm{for} \ |x_\mathrm{p}(t)| \leq R_\star
\end{cases}
\end{equation}

\noindent
\\
where $x_\mathrm{p}(t)$ is the position of the planet, and

\begin{equation}
A_\mathrm{mask}  = \frac{2}{\pi}  \sum_{\mathrm{s}=1}^n \left( \frac{R_\mathrm{s}}{R_\mathrm{p}} \right)^2 \left[ \arccos\left( \frac{-R_\mathrm{p}}{a_{\mathrm{ps}}} \right) - \arccos\left( \frac{+R_\mathrm{s}}{a_{\mathrm{ps}}} \right) \right]
\end{equation}

\noindent
\\
compensates for those parts of the probability functions that do not contribute to the OSE because of planet-moon eclipses. This masking can only occur during the planetary transit when $|x_\mathrm{p}(t)| \leq R_\star$. Note that partial planet-moon eclipses as well as moon-moon eclipses are ignored \citep[but treated by][]{2011MNRAS.416..689K}.

In this model, the ingress and egress of the moons are neglected, which is appropriate because even the largest moons that can possibly form within the circumplanetary disk around a 10-Jupiter-mass planet are predicted to have masses around ten times that of Ganymede, or $\approx0.25\,M_\oplus$ \citep{2006Natur.441..834C}, and even if they are water-rich their radii will be $<R_\oplus$. The effect of a moon's radial extension on the duration of the OSE will thus be $<R_\oplus/(5\,R_\mathrm{J})\approx1.8\,\%$ for a sub-Earth-sized moon at a planet-moon orbital distance of 5 Jupiter radii ($R_\mathrm{J}$) and $<0.5\,R_\oplus/(10\,R_\mathrm{J})\approx0.46\,\%$ for a Mars-sized moon with a semi-major axis of $10\,R_\mathrm{J}$.

\subsubsection{Numerical Simulations of the Photometric OSE}
\label{sub:OSEnumerical}

To simulate a light curve that contains a photometric OSE, I construct a hypothetical three-satellite system that is similar in scale to the three innermost moons of the Galilean system. The planet is assumed to have the 10-fold mass of Jupiter and a radius $R_\mathrm{p}=1.05$ Jupiter radii ($R_\mathrm{J}$). According to the \citet{2006Natur.441..834C} model, I scale the satellite masses as $M_1=10\,M_\mathrm{I}$, $M_2=10\,M_\mathrm{E}$, and $M_3=10\,M_\mathrm{G}$, with the indices I, E, and G referring to Io, Europa, and Ganymede, respectively. I derive the moon radii of this scaled-up system as per the \citet{Fortney2007} structure models for icy/rocky planets by assuming ice-to-mass fractions (imf) similar to those observed in the Jovian system, that is, $\mathrm{imf}_1=0.02$, $\mathrm{imf}_2=0.08$, and $\mathrm{imf}_3=0.45$ \citep{2009euro.book...59C}. The model then yields $R_1=0.62\,R_\oplus$, $R_2=0.52\,R_\oplus$, and $R_3=0.86\,R_\oplus$. The star is assumed to be a K star 0.7 times the mass of the Sun ($M_\odot$) with a radius ($R_\star$) of 0.64 solar radii ($R_\odot$) \citep[for Sun-like metallicity at an age of 1\,Gyr, following][]{2012MNRAS.427..127B}, and the planet-satellite system is placed into the center of the stellar HZ at 0.56\,AU \citep[derived from the model of][]{2013ApJ...765..131K}. The impact parameter is $b=0$, the moons' orbits are all in the orbital plane of the planet-moon barycenter around the star, and the satellites' orbital semi-major axes around the planet are $a_{\mathrm{p}1}=a_{\mathrm{JI}}R_\mathrm{p}/R_\mathrm{J}$, $a_{\mathrm{p}2}=a_{\mathrm{JE}}R_\mathrm{p}/R_\mathrm{J}$, and $a_{\mathrm{p}3}=a_{\mathrm{JG}}R_\mathrm{p}/R_\mathrm{J}$ for the innermost, the central, and the outermost satellite, respectively. The values of $a_{\mathrm{JI}}$, $a_{\mathrm{JE}}$, and $a_{\mathrm{JG}}$ correspond to the semi-major axes of Io, Europa, and Ganymede around Jupiter, respectively, and consequently this system is similar to the one shown in Figure~\ref{fig:ProbDens}.

\begin{figure}[t]
  \centering
  \scalebox{0.53}{\includegraphics{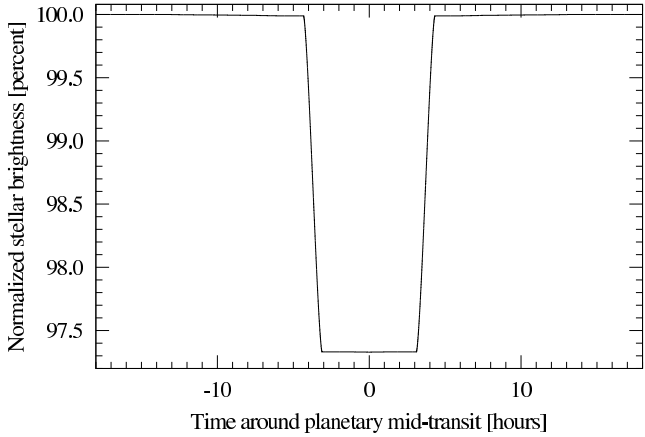}}
  \caption{Simulated transit light curve of a hypothetical three-satellite system around a giant planet of 1.05 Jupiter radii and 10 Jupiter masses transiting a $0.64\,R_\odot$ K star in the HZ. On this scale, the photometric OSE is barely visible as a small decrease in stellar brightness just before planetary ingress and as a small delay in reaching $100$\,\% of stellar brightness after planetary egress. Stellar limb darkening is neglected.}
  \label{fig:transit}
\end{figure}

\begin{figure*}[t]
  \centering
  \scalebox{0.455}{\includegraphics{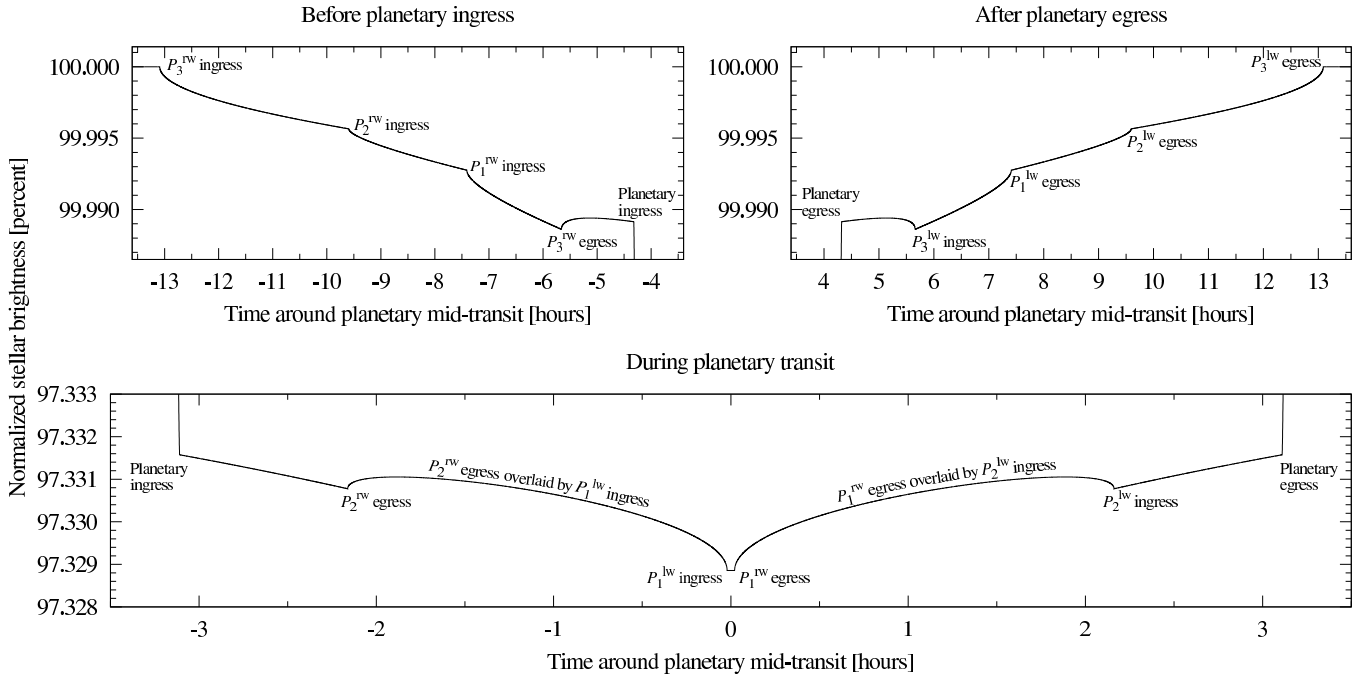}}
  \caption{Photometric OSEs of a hypothetical three-satellite system around a super-Jovian planet transiting a $0.64\,R_\odot$ K star in the HZ (zoom into Figure~\ref{fig:transit}). The radii of the outermost (the 3rd), central (2nd), and innermost (1st) satellites are $R_3 = 0.86\,R_\oplus$, $R_2 = 0.52\,R_\oplus$, and $R_1 = 0.62\,R_\oplus$, following planetary structure models \citep{Fortney2007} and assuming ice-to-mass fractions of the Galilean moons. Due to the smaller range of orbits over which the probability function $P(a_{\mathrm{p}1})$ of the innermost satellite is spread, its dip in this averaged stellar light curve is deeper than the brightness decrease induced by the outermost satellite, although the innermost satellite is smaller ($R_3>R_1$). An arbitrarily large number of transits has been averaged to obtain this curve, and no noise has been added.}
  \label{fig:zoom}
\end{figure*}

\begin{figure*}[t]
  \centering
  \scalebox{0.44}{\includegraphics{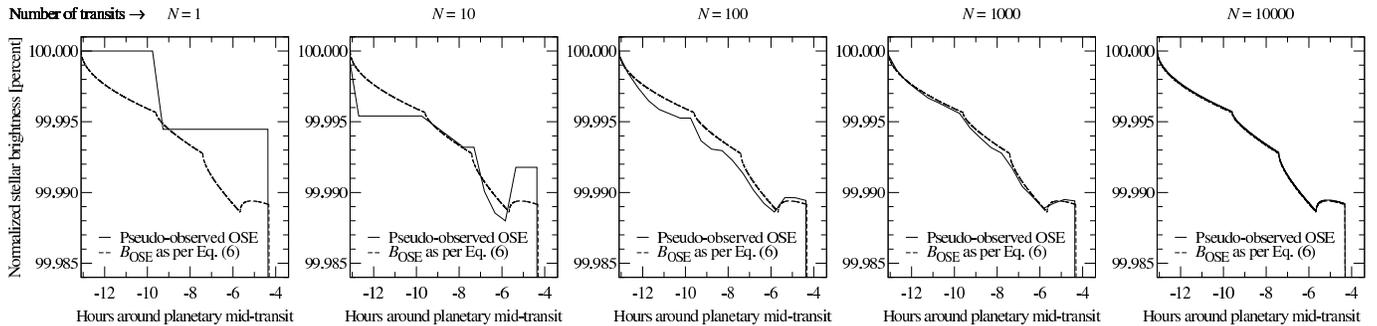}}\\
  \vspace{0.2cm}
  \caption{Emergence of the photometric OSE during the transit ingress of a hypothetical three-satellite system for an increasing number of averaged transits $N$ (zoom into upper left panel of Figure~\ref{fig:zoom}). Solid lines show the averaged noiseless light curves, while dashed curves illustrate the combined photometric OSEs of the three satellites as per Equation~(\ref{eq:OSE_multi}).}
  \label{fig:evol}
\end{figure*}

Figure~\ref{fig:transit} shows the transit light curve of this system, averaged over an arbitrary number of transits with infinitely small time resolution, excluding any sources of noise, and neglecting effects of stellar limb darkening. Individual transits are modeled numerically by assuming a random orbital positioning of the moons during each transit. In reality, a moon's relative position to the planet during a transit is determined by the initial conditions, say at the beginning of some initial transit, as well as by the orbital periods both around the star and around the planet-moon barycenter. I here only assume that the ratio of these periods is not a low value integer. Then from a statistical point of view, positions during consecutive transits can be considered randomized.

In the simulations shown in Figure~\ref{fig:transit}, the analytic description of Equation~(\ref{eq:OSE_multi}) is not yet applied -- this pseudo phase-folded light curve is a randomized, purely numerical simulation. If one of the moons turns out to be in front of or behind the planet, as seen by the observer, then it does not cause a flux decrease in the light curve. On this scale, the combined OSE of the three satellites is hardly visible against the planetary transit because the depth of the satellites' features scale as $(\approx0.64\,R_\oplus/(0.64\,R_\odot))^2=10^{-4}$, while the planet causes a depth of about $2.7\,\%$, decreasing stellar brightness to roughly $97.3\,\%$.

Figure~\ref{fig:zoom} zooms into three parts of this light curve. The upper left panel highlights the respective ingresses of the three satellites. Each of the three impressions is caused by the right wing of the probability function of one of the moons, with the outermost moon (the third moon counting outwards from the planetary center) entering first, about 13 hr before the planetary mid-transit, the second moon following (ingress at about $-10$ hr), and the inner, or first, moon succeeding at roughly $-7.5$ hr along the abscissa. Each individual OSE ingress corresponds to the phase between moments (1) and (2) in Figure~\ref{fig:OSE} and is tied to the photometric OSE of the preceding moon or moons. As the moons' probability functions enter the stellar disk, their OSE signals add up. At about $-4.5$ hr, the planetary ingress begins and causes a steep decrease in stellar brightness. The slight increase of the curve between $-5.5$ and $-4.5$ hr is caused by the right wing ($P_3^\mathrm{rw}$) of the outermost moon's probability function leaving the stellar disk even before the planet enters. This indicates that the projected orbital separation of the outermost satellite is larger than the radius of the stellar disk.

The upper right panel shows the egress of the planet and the three-satellite system from the stellar disk, which appears as a mirror-inverted version of the upper left panel. In this panel, the left wings of the satellites' probability functions that leave the stellar disk replace the right wings that enter the disk from the left panel.

The wide lower panel of Figure~\ref{fig:zoom} illustrates the moons' photometric OSEs at the bottom of the planetary transit trough. Chronologically, this phase of the transit is between the ingress (upper left) and egress (upper right) phases of the planet-satellite system. The downtrend between about $-3$ and $-2.2$ hr visualizes the continued ingresses of $P_1^\mathrm{rw}$, $P_2^\mathrm{rw}$, and $P_3^\mathrm{rw}$ from the upper left panel. What is more, between about $-2.2$ hr and the center of the planetary transit curve, we witness an interplay of the satellites' $P_\mathrm{s}(x)$ entering and leaving the stellar disk. In particular, the long curved feature between approximately $-2.2$ and $0$ hr visualizes the egress of the right wing of the central moon ($P_2^\mathrm{rw}$) overlaid by the ingress of increasingly high sampling frequencies in the left wing of the central moon ($P_1^\mathrm{lw}$). The fact that the end of the ingress of $P_1^\mathrm{lw}$ and the beginning of the egress of $P_1^\mathrm{rw}$ almost coincide at planetary mid-transit means that the width of the projected semi-major axis of the innermost satellite equals almost exactly the stellar radius.

\subsubsection{Emergence of the Photometric OSE in Transit Light Curves}
\label{sub:emergence}

Assuming the moons do not change their positions relative to the planet during individual transits, then their apparent separations are discrete in the sense that they are defined by one value. Only if numerous transits are observed and averaged will the moons' photometric OSEs appear, because the sampling frequencies, or the underlying density distributions $P_\mathrm{s}(x)$, will take shape. Dropping the assumption of a fixed planet-satellite separation during transit, each individual transit will cause a dynamical imprint in the light curve. However, the photometric OSE will converge to the same analytical expression as given by Equation~(\ref{eq:OSE_multi}) after a large number of transits.

In Figure~\ref{fig:evol}, I demonstrate the emergence of the photometric OSE in the hypothetical three-satellite system for an increasing number of transits $N$. Each panel shows the same time interval around planetary mid-transit as the upper left panel in Figure~\ref{fig:zoom}, that is, the ingress of the right wings of the probability distributions. The solid line shows the averaged transit light curve coming from my randomized transit simulations, and the dashed line shows the predicted OSE signal as per Equation~(\ref{eq:OSE_multi}). In order to draw the dashed line, I make use of the \textit{known} satellite radii $R_\mathrm{s}$ ($\mathrm{s}=1,2,3$), the planet-satellite semi-major axes $a_{\mathrm{ps}}$, the stellar radius, the planetary radius, and the orbital velocity of the transiting planet-moon system $v_\mathrm{orb}$. The dashed curve is thus no fit to the simulations but a prediction for this particular exomoon system.

In the left-most panel, after the first transit of the planet-satellite system in front of the star ($N=1$), only one of the three moons shows a transit, shortly after $-10$ hr. Referring to the picture given in Figure~\ref{fig:OSE}, this moon appears to the right of the planet and enters the stellar disk before the planet does. Examination of this moon's transit depth of $6{\times}10^{-5}$ reveals the second moon as the originator, because $(R_2/R_\star)^2\approx6{\times}10^{-5}$. As an increasing number of transits is collected in the second-to-left and the center panel, the photometric OSEs of the three-satellite system emerge. Obviously, between $N=10$ and $N=100$, a major improvement of the OSE signal strength occurs, suggesting that at least a few dozen transits are necessary to characterize this system. After $N=1000$ transits, the noiseless averaged OSE curve becomes indistinguishable from the predicted function.

Whatever the precise value of the critical number of transits ($N_\mathrm{OSE}$) necessary to recover the satellite system from the solid curve in Figure~\ref{fig:evol}, it is a principled threshold imposed by the very nature of the OSE. Noise added during real observations increases the number of transits that is required to characterize the system to a value $N_\mathrm{obs}$, and the relation is deemed to be $N_\mathrm{obs}~{\geq}~N_\mathrm{OSE}$. To estimate realistic values for $N_\mathrm{obs}$, it is necessary to simulate noisy pseudo-observed data and try to recover the input systems. Section~\ref{sec:detections} is devoted to this task.

\subsection{OSE of Exomoon-induced Transit Duration Variations (TDV-OSE)}
\label{subsec:TDV-OSE}

Consider a single massive exomoon orbiting a planet. As a result of the two bodies' motion around their common barycenter, the planet's tangential velocity component with respect to the observer is different during each transit, and hence the duration of the planetary transit shows deviations from the mean duration during each transit \citep{2009MNRAS.392..181K}. While each individual planet-moon transit has its individual TDV offset, all TDV observations combined will reveal what I refer to as a TDV-OSE. In the reference system shown in Figure~\ref{fig:geometry}, the planet's velocity component is projected onto the $x$-axis and is given by $\dot{x}_\mathrm{p} = {\omega}r_\mathrm{p}\sin(\varphi)$, with $\omega~=~2\pi/P_\mathrm{ps}$ as the planet-satellite orbital frequency and $P_\mathrm{ps}$ as the orbital period. The probability for a single planetary transit to show a certain projected velocity around the planet-satellite barycenter is then given by the sampling frequency

\begin{align} \label{eq:TDVOSE} \nonumber
P_\mathrm{p}^\mathrm{TDV}(\dot{x}_\mathrm{p}) &\propto \frac{\mathrm{d}\varphi}{\mathrm{d}\dot{x}_\mathrm{p}} = \frac{\mathrm{d}}{\mathrm{d}\dot{x}_\mathrm{p}} \arcsin\left(\frac{\dot{x}_\mathrm{p}}{{\omega}r_\mathrm{p}}\right) \\
        &= \frac{1}{{{\omega}r_\mathrm{p}}Ê\wu{1-\left(\frac{\displaystyle \dot{x}_\mathrm{p}}{\displaystyle {\omega}r_\mathrm{p} }\right)^2 }}  \ \ .
\end{align}

\noindent
This distribution describes the fraction of randomly sampled angles $\varphi$ that lies within an infinitesimal velocity interval $\mathrm{d}\dot{x}_\mathrm{p}$ of the planet projected onto the $x$-axis, whereas $P_\mathrm{s}(x)$ in Equation~(\ref{eq:OSE_basic}) measures how much of an infinitely small orbital path element of a satellite lies in a projected planet-moon distance interval. In analogy to the normalization of the latter position probability in Eq.~(\ref{eq:norm}), the integral over $P_\mathrm{p}^\mathrm{TDV}(\dot{x}_\mathrm{p})$ must equal 1 between $-{\omega}r_\mathrm{p}$ and $+{\omega}r_\mathrm{p}$, which yields

\begin{equation}\label{eq:ProbDens_xdot}
P_\mathrm{p}^\mathrm{TDV}(\dot{x}_\mathrm{p}) = \frac{1}{{\pi\omega}r_\mathrm{p}\wu{1-\left(\frac{\displaystyle \dot{x}_\mathrm{p}}{\displaystyle {\omega}r_\mathrm{p} }\right)^2 }} \ \
\end{equation}

\noindent
and has a shape similar to the functions shown in Figure~\ref{fig:ProbDens}. Consequently, as the planet's velocity wobble induced by a single moon distributes within the interval $-{\omega}r_\mathrm{p}\leq\dot{x}_\mathrm{p}\leq{\omega}r_\mathrm{p}$ as per Equation~(\ref{eq:ProbDens_xdot}), planetary TDVs will also distribute according to an OSE.

Assuming circular orbits, the spread of this TDV distribution is determined by the peak-to-peak TDV amplitude

\begin{equation}\label{eq:DeltaTDV}
\Delta_\mathrm{TDV} = 2 \ t_\mathrm{T} \times \wu{\frac{a_{{\star}\mathrm{p}}}{a_\mathrm{ps}}} \wu{ \frac{M_\mathrm{s}^2}{M_\mathrm{b}(M_\mathrm{b}+M_\star)} } \ \ ,
\end{equation}

\noindent
where $t_\mathrm{T}$ is the duration of the transit between first and fourth contact \citep{2009MNRAS.392..181K}. Then the TDV-OSE distribution is given by

\begin{equation}\label{eq:ProbDens_TDV}
P_\mathrm{p}^\mathrm{TDV}(t) = \frac{1}{{\pi\Delta_\mathrm{TDV}} \wu{1-\left(\frac{\displaystyle t}{\displaystyle \Delta_\mathrm{TDV} }\right)^2 } } \ \ ,
\end{equation}

\noindent
where $t$ is time. With $M_\star{\gg}M_\mathrm{p}$ and $M_\mathrm{p}{\gg}M_\mathrm{s}$ the right-most square root in Equation~(\ref{eq:DeltaTDV}) simplifies to $\wu{M_\mathrm{s}^2/(M_\mathrm{p}M_\star)}$. $M_\star$ can be determined via spectral classification, $M_\mathrm{p}$ may be accessible via radial velocity measurements, $t_\mathrm{T}$ is readily available from the light curve, $a_{\star\mathrm{p}}$ can be inferred by Kepler's third law, and $a_\mathrm{ps}$ can be measured by the photometric OSE. With $M_\mathrm{s}$ as the remaining free parameter, a fit of Equation~(\ref{eq:ProbDens_TDV}) to the moon-induced planetary TDV distribution gives a direct measurement of the satellite mass. But note the limited applicability of the TTV and TDV methods for short-period moons \citep[Section 6.3.8 in][]{2011tepm.book.....K}.

\subsection{OSE of Exomoon-induced Transit Timing Variations (TTV-OSE)}
\label{subsec:TTV-OSE}

There is, of course, also an OSE for the distribution of the TTVs. In circularÊ\, one-satellite systems, the amplitude of this TTV-OSE distribution is given by

\begin{equation}
\Delta_\mathrm{TTV} = 2 \ \times \frac{a_\mathrm{ps}M_\mathrm{s}}{M_\mathrm{b}\wu{a_\mathrm{{\star}p}G(M_\star+M_\mathrm{b})}} \ \
\end{equation}

\noindent
\citep{1999A&AS..134..553S,2009MNRAS.392..181K}, and the distribution itself is given by

\begin{equation}\label{eq:ProbDens_TTV}
P_\mathrm{p}^\mathrm{TTV}(t) = \frac{1}{{\pi\Delta_\mathrm{TTV}} \wu{1-\left(\frac{\displaystyle t}{\displaystyle \Delta_\mathrm{TTV} }\right)^2 } } \ \ .
\end{equation}

\noindent
With all the parameters on the right-hand side of Eq.~(\ref{eq:ProbDens_TTV}) known from radial velocity measurements, the photometric OSE and TDV-OSE, this third manifestation of the OSE can be used to further improve the confidence of any exomoon detection, as it needs to be consistent with all three OSE observations.\\

\section{Recovering exomoon-induced photometric OSE with \textit{Kepler}}
\label{sec:detections}

To assess the prospects of measuring the photometric OSE as described in Section~\ref{subsec:OSEanalytic}, I generate a suite of pseudo-observed averaged light curves for a range of given star-planet-satellite systems and try to detect the injected exomoon signal. Following \citet{2009MNRAS.400..398K}, I assume that the out-of-transit baseline is known to a sufficiently high degree, which is an adequate assumption for \textit{Kepler} high-quality photometry with large amounts of out-of-transit data. My simulations also imply that red noise has been corrected for, which can be a time-consuming part of the data reduction. Stellar limb darkening is neglected, but this does not alter the photometric OSE substantially (Heller et al., in preparation).

\subsection{Noise and Binning of Phase-Folded Light Curves}
\label{sub:noise}

\begin{figure}[t]
  \centering
  \scalebox{0.44}{\includegraphics{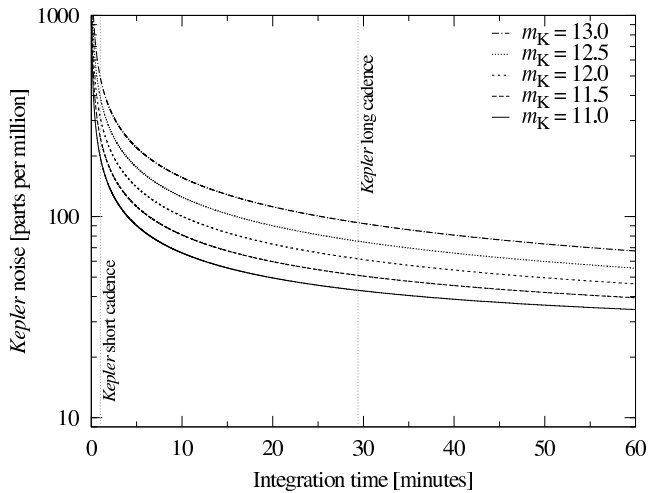}}
  \caption{Simulated \textit{Kepler} noise as per Equation~(\ref{eq:Keplernoise}), used for the synthesis of pseudo-observed transit light curves.}
  \label{fig:Keplernoise}
\end{figure}

\begin{figure*}[t]
  \centering
  \scalebox{0.455}{\includegraphics{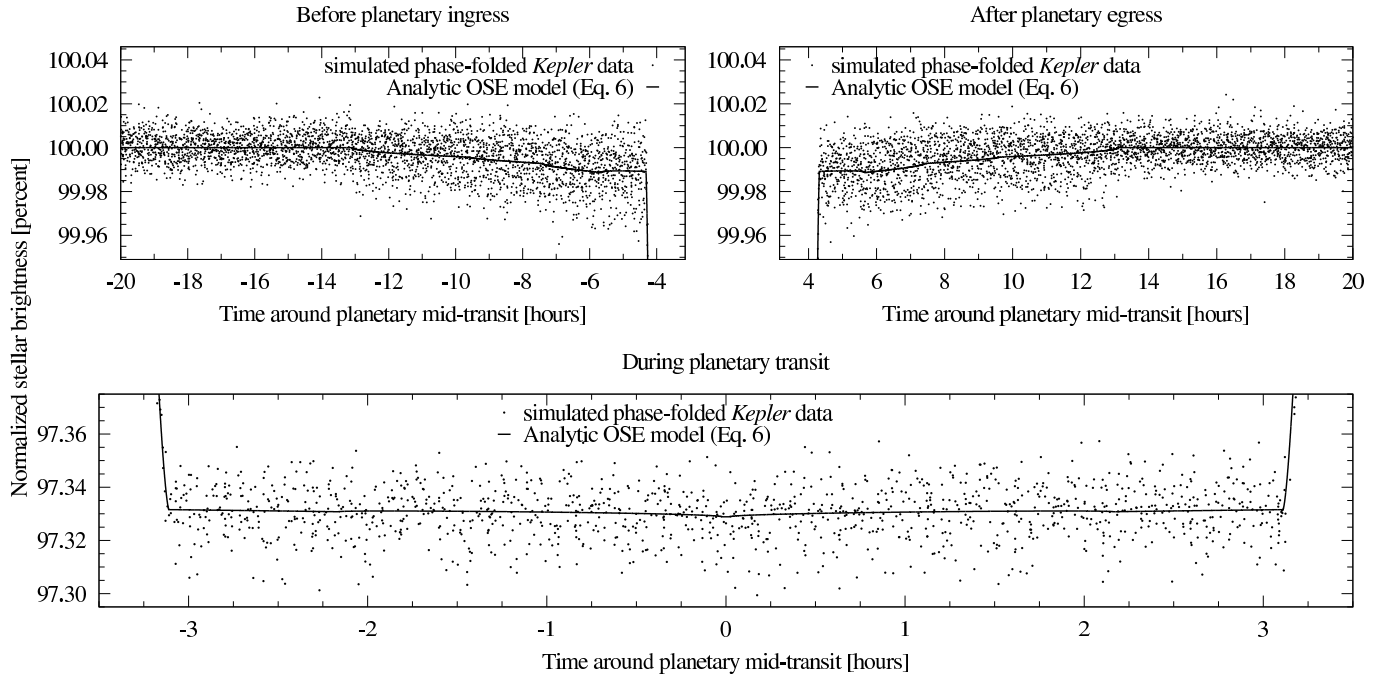}}
  \caption{Photometric OSEs after 100 transits in simulated \textit{Kepler} observations of a hypothetical three-satellite system around a Jupiter-sized planet 10 times the mass of Jupiter, transiting a $0.64\,R_\odot$ K star in the HZ (same as in Figure~\ref{fig:zoom}). Noise is simulated after Equation~(\ref{eq:Keplernoise}). The scale of the ordinate is much wider than in Figure~\ref{fig:zoom}. Although the OSE seems invisible at the bottom of the transit light curve (bottom panel) due to the noise, the three moons together mask a measurable amount of stellar light that triggers the fit.}
  \label{fig:Keplerobs}
\end{figure*}

\begin{figure*}[t]
  \centering
  \scalebox{0.455}{\includegraphics{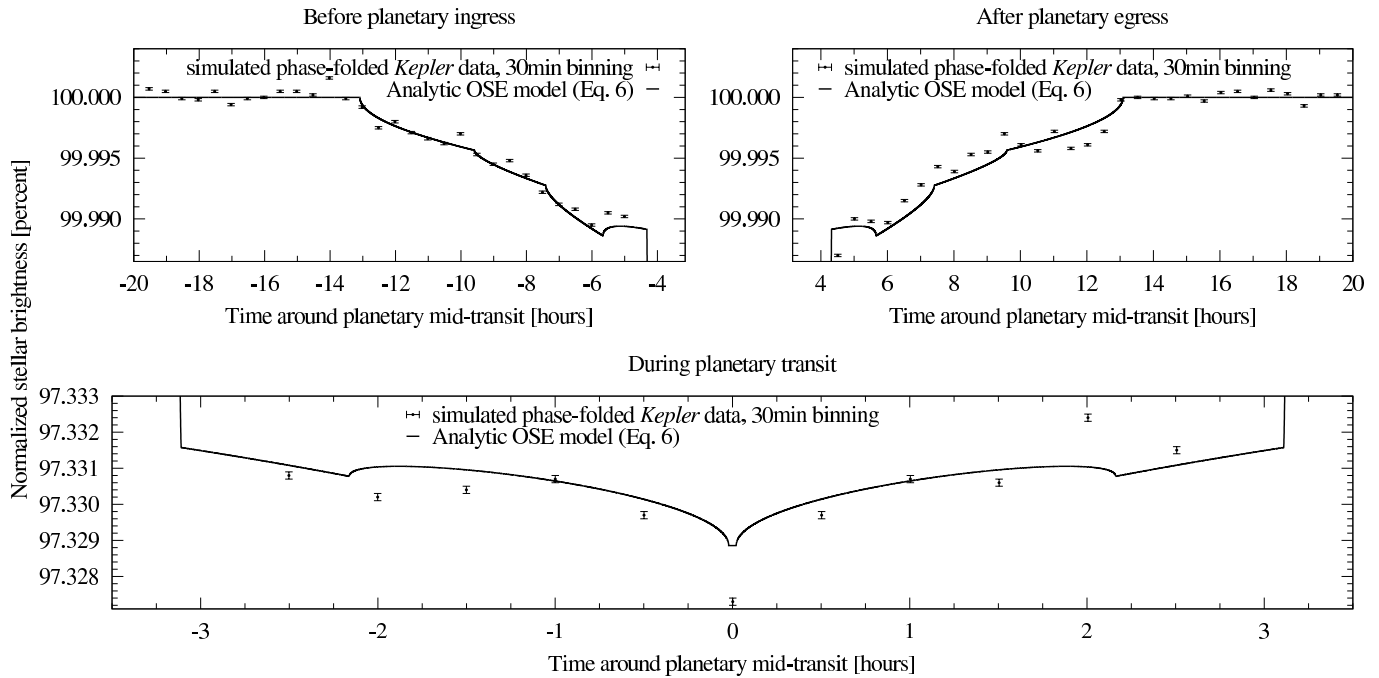}}
  \caption{Same as Figure~\ref{fig:Keplerobs} but binned to intervals of 30 min. While the OSE of the three-satellite system clearly emerges in the ingress and egress parts of the probability functions $P_\mathrm{s}(x)$ (upper two panels), it remains hardly visible at the bottom of the transit curve. Yet, the combined stellar transits of the moons still influence the depth of this light curve trough.}
  \label{fig:Keplerbin}
\end{figure*}

\begin{figure*}[t]
  \centering
  \scalebox{0.455}{\includegraphics{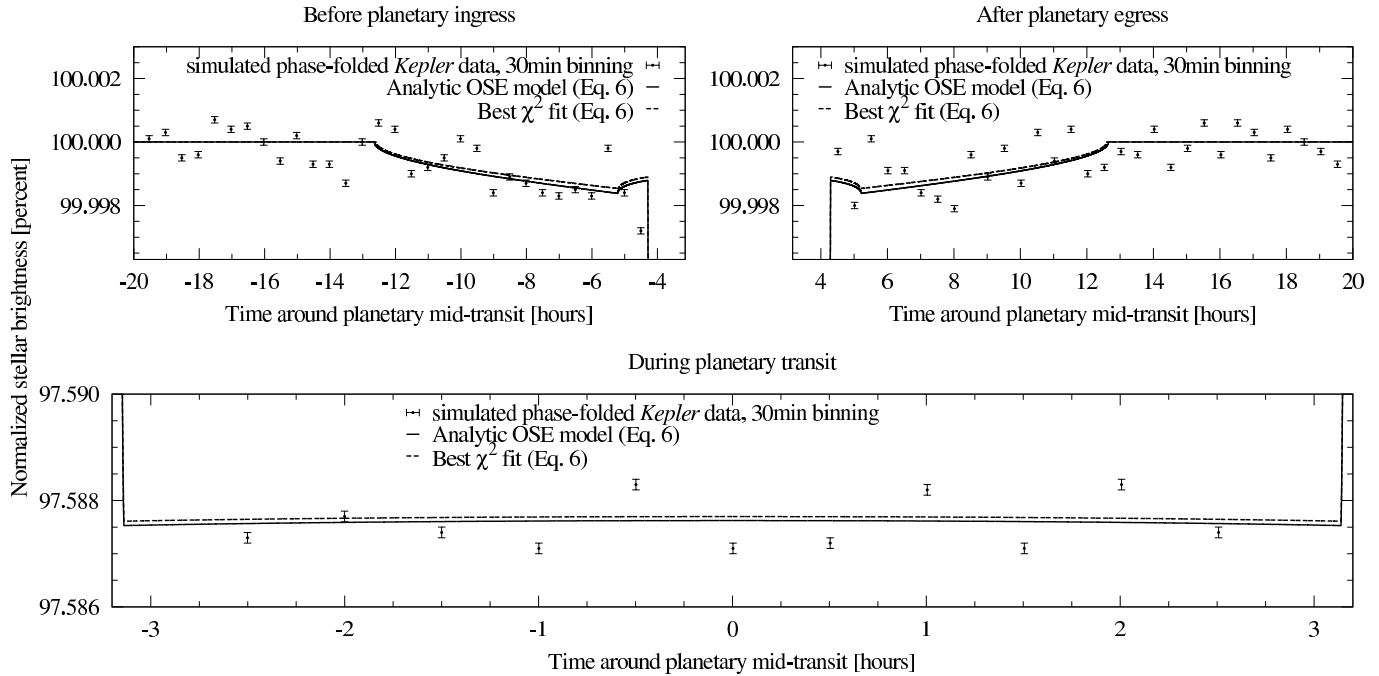}}
  \caption{$\chi^2$ fit of the analytical OSE model via Equation~(\ref{eq:OSE_multi}) (dashed lines) to the binned, simulated \textit{Kepler} data of a one-satellite system (data points) after $N=100$ transits. The solid line shows the OSE light curve for the \textit{known} input system. The moon is similar to Ganymede in terms of mass ($0.025\,M_\oplus$), radius ($0.42\,R_\oplus$), and planet-moon distance ($15.47\,R_\mathrm{p}$). In this simulation, the $\chi^2$ fit yields a moon radius $R_1=0.4\,\pm0.0049\,R_\oplus$ and planet-moon semi-major axis $a_\mathrm{p1}=15.4\,\pm0.0487\,R_\mathrm{p}$.}
  \label{fig:Keplerbinfit}
\end{figure*}

\begin{figure*}[t]
  \centering
  \scalebox{0.333}{\includegraphics{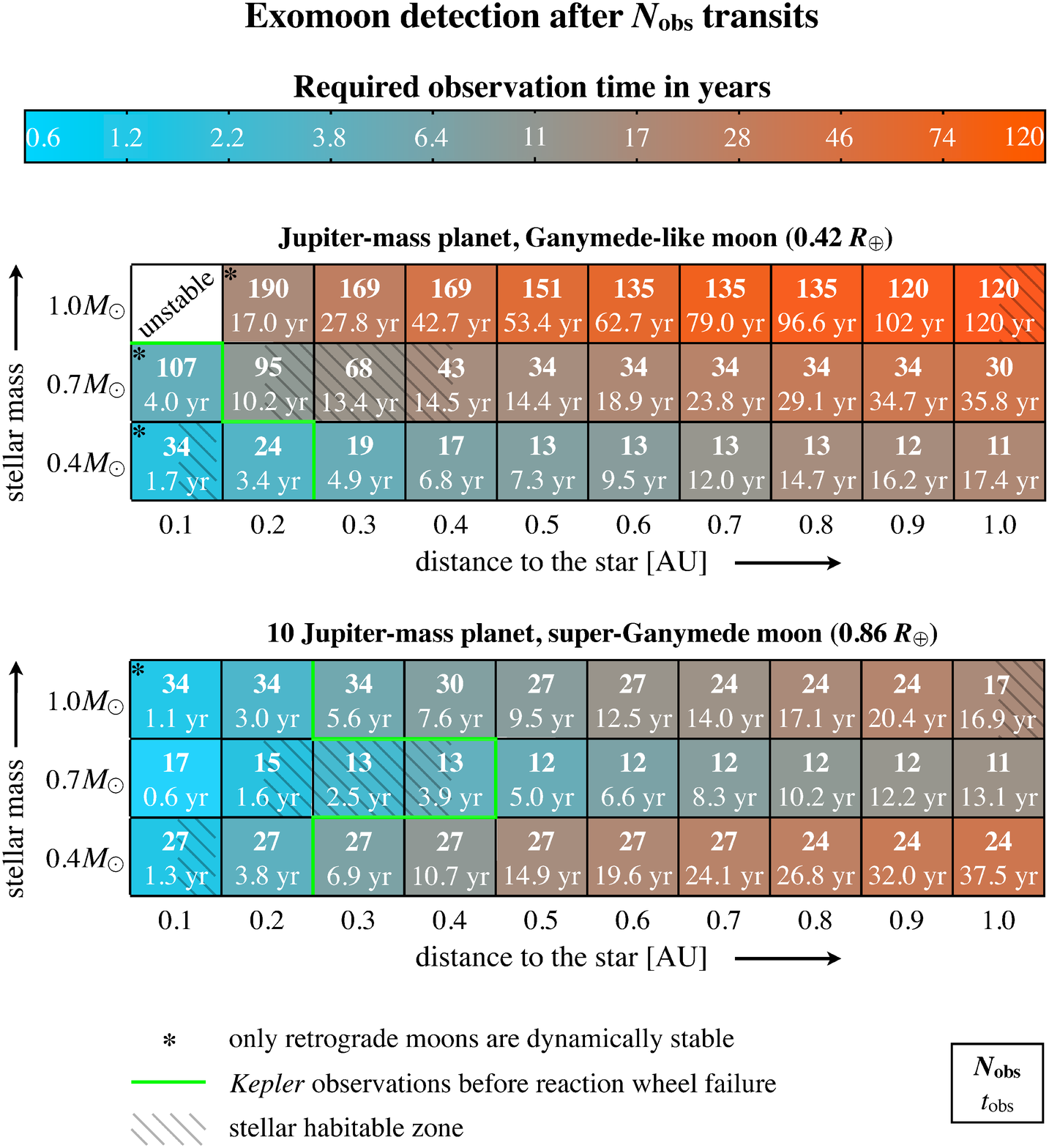}}
  \caption{Number of transits required to detect the photometric OSE ($N_\mathrm{obs}$, bold numbers) and equivalent observation time of a one-satellite system around a Jupiter-sized planet ( upper panel) and a 10 Jupiter-mass planet (lower panel). The three rows correspond to a 0.4, 0.7, and $1\,M_\odot$-mass host star, respectively, while the columns depict the semi-major axis of the planet-moon binary around the star. Striped areas indicate the stellar HZ, and green cell borders embrace the observation cycle that has been covered with \textit{Kepler} before its reaction wheel failure. In all simulations, the moon is assumed to orbit in a Ganymede-wide orbit around the planet, that is, at about $15\,R_\mathrm{p}$.}
  \label{fig:Nobs}
\end{figure*}

To simulate a \textit{Kepler}-class photometry, \,ÊI induce stellar brightness variations $\sigma_\star$, detector noise $\sigma_\mathrm{d}$, a quarter-to-quarter noise component $\sigma_\mathrm{q}$, and shot noise $\sigma_\mathrm{s}$ into the raw light curves that contain a moon-induced photometric OSE. The total noise is then given by

\begin{equation} \label{eq:Keplernoise}
\sigma_\mathrm{K} = \wu{\sigma_\star^2 + \sigma_\mathrm{d}^2 + \sigma_\mathrm{q}^2 + \frac{1}{\Gamma_\mathrm{ph}t_\mathrm{int}}} \ \ ,
\end{equation}

\noindent
with $\sigma_\star=19.5$\,ppm, $\sigma_\mathrm{D}=10.8$\,ppm, $\sigma_\mathrm{Q}=7.8$\,ppm \citep{2011ApJS..197....6G},

\begin{equation}
\Gamma_\mathrm{ph}=6.3\times10^8\,\mathrm{hr}^{-1}\times10^{-0.4(m_\mathrm{K} - 12)}
\end{equation}

\noindent
as \textit{Kepler}'s photon count rate \citep{2009MNRAS.400..398K}, and $t_\mathrm{int}$ as the integration time. The stellar noise of 19.5\,ppm is typical for a G-type star, while K and M dwarfs tend to show more intrinsic noise. Figure~\ref{fig:Keplernoise} visualizes the decrease of \textit{Kepler} noise as a function of integration time and for five different stellar magnitudes. The \textit{Kepler} short cadence and long cadence integration times at 1 and 29.4\,min are indicated with vertical lines, respectively.

Figure~\ref{fig:Keplerobs} shows the pseudo phase-folded transit light curve of the hypothetical three-satellite system described in Section~\ref{sub:OSEnumerical} after 100 transits. The K star is chosen to have a \textit{Kepler} magnitude $m_\mathrm{K}~=~12$. I simulate each individual transit and virtually ``observe'' the stellar brightness every 29.4\,min, corresponding to the \textit{Kepler} \,Êlong-cadence mode, and add Gaussian noise following Equation~(\ref{eq:Keplernoise}). For each transit, I introduce a random timing offset to the onset of observations, so that each transit light curve samples different parts of the transit. Each individual light curve is normalized to 1 and added to the total pseudo phase-folded light curve. This pseudo phase-folded light curve is again normalized to 1. While dots in Figure~\ref{fig:Keplerobs} visualize my simulated \textit{Kepler} measurements, the solid line depicts the analytic prediction of the OSE following Equation~(\ref{eq:OSE_multi}). Even in this noisy data, the OSE is readily visible with the unaided eye shortly before and after the planetary transit (upper two panels), suggesting that less than 100 transits are necessary to discover -- yet maybe not to unambiguously characterize -- extrasolar multiple moon systems with the photometric OSE. At the bottom of the pseudo phase-folded transit light curve, however, the OSE remains hardly noticeable (lower panel).

In Figure~\ref{fig:Keplerbin}, I show the same simulated data, but now binned to intervals of 30\,min (see Appendix~\ref{sec:binning}). Before and after planetary ingress, the OSE now becomes strongly apparent. Note that the standard deviations are substantially smaller than the scatter around the analytic model. This is not due to observational noise, but due to the discrete sampling of the transits. This scatter from the model decreases for an increasing number of transits, $N$ (see Figure~\ref{fig:evol}). I also tried binning the short cadence \textit{Kepler} data and used binning intervals of 10, 30, and 60\,min, of which the 30\,min binning of long cadence observations showed the most reliable results, at least for this particular star-planet-moon system. With a 60\,min binning, the individual ingress and egress of the three moons are poorly sampled, while the 10\,min sampling shows too much of a sampling scatter around the model.

\subsection{Recovery of Injected Exomoon OSE Signals}
\label{subsec:recovery}

Next, I evaluate the odds of characterizing extrasolar moons with the photometricÊ\, OSE. Although multi-satellite systems can, in principle, be detected and characterized by the analytical OSE model (Equation~(\ref{eq:OSE_multi})), I focus on a one-satellite system. I simulate a range of transits of a single exomoon orbiting a Jupiter- and 10-Jupiter-mass planet. In the former case, the moon is a Ganymede analog of $0.42\,R_\oplus$, in the latter case the moon's mass is scaled by a factor of ten and its radius of $0.86\,R_\oplus$ is derived from the \citet{Fortney2007} structure models using $\mathrm{imf}_1=0.45$. This moon corresponds to the outermost (or third) moon in the system simulated in Section~\ref{subsec:OSEsignatures}. I study the detectability of these two moons around these two planets orbiting three different stars at 12th magnitude: a Sun-like star, a $0.7\,M_\odot$ K star as considered in the previous sections, and an $0.4\,M_\odot$-mass M dwarf with a radius of $0.36\,R_\odot$ \citep[for solar metallicity at an age of 1\,Gyr, derived from][]{2012MNRAS.427..127B}.

I start to simulate the pseudo phase-folded \textit{Kepler} light curve of each of these hypothetical systems after $N=5$ transits and fit the binned data with a $\chi^2$ minimization of the analytical model given by Equation~(\ref{eq:OSE_multi}), where the two free parameters are the satellite radius ($R_1$) and the orbital semi-major axis between the moon and the planet ($a_{\mathrm{p}1}$) (see Appendix~\ref{sec:fitting}). This procedure is performed 100 times for a given $N$, and if both $R_1$ and $a_{\mathrm{p}1}$ are recovered with an error of less than $10\,\%$ in at least 68 of the 100 runs (corresponding to a $1\,\sigma$ confidence for the recovery rate), then the number of transits is stored as $N_\mathrm{obs}$. If the satellite cannot be recovered under these boundary conditions, then the number of transits $N$ is increased by an amount ${\Delta}N=\lfloor10^{\log_{10}(N) + 0.05}\rceil$,\footnote{The notation ${\lfloor}x{\rceil}$ denotes a rounding of the real number $x$ to the next integer.} representing the series $N\in\{5, 6, 7, ..., 48, 54, 61, ..., 151,169, 190, ... \}$, and I repeat my fit of the analytical model to 100 pseudo phase-folded light curves.

Figure~\ref{fig:Keplerbinfit} presents an example light curve of the Jupiter-mass planet and its Ganymede-like satellite in the K star toy system after $N=100$ transits. The data points show the simulated \textit{Kepler} observations, the solid line corresponds to the predicted light curve of the known system, and the dashed line indicates the best fit model. In this example, the best-fit radius of the moon of $0.4\,\pm0.0049\,R_\oplus$ is very close to the input value of $0.42\,R_\oplus$, and similarly the fit of the planet-moon orbital distance of $15.4\,\pm0.0487\,R_\mathrm{p}$ almost matches the input value of $15.47\,R_\mathrm{p}$. I consider the formal $1\sigma$ uncertainties of the $\chi^2$ fitting as physically unrealistic, which is why I repeat the fitting procedure 100 times to get more robust estimates of $N_\mathrm{obs}$.

\section{Results and predictions}
\label{sec:results}

Figure~\ref{fig:Nobs} shows the outcome of the data fitting. The upper table refers to transits of a Jupiter-sized host planet with a Ganymede-like moon, and the lower table lists the results for the super-Jovian planet with a super-Ganymede exomoon. Abscissae and ordinates of these charts indicate distance to the star and stellar mass, respectively, while the entries show $N_\mathrm{obs}$ as well as the equivalent observational time $t_\mathrm{obs}$, computed as $N_\mathrm{obs}$ times the orbital period around the star. Assuming circular orbits, I use the semi-analytic model of \citet{2006MNRAS.373.1227D} to test all systems for orbital stability. While a Jupiter-mass planet 0.1\,AU from a Sun-like star cannot hold a moon in a Ganymede-wide orbit, a few other configurations only allow the moon to be stable in retrograde orbital motion. The latter cases are labeled with an asterisk. Shaded regions indicate the locations of the respective stellar HZ following \citet{2013ApJ...765..131K}. Green cell borders demarcate the observation cycle of the \textit{Kepler} space telescope that has been covered before the satellite's reaction wheel failure. Photometric OSEs of exomoons within these boundaries could be detectable in the available \textit{Kepler} data of photometrically quiet stars.

Inspecting the upper panel for the Jupiter-Ganymede duet, three trends readily appear. First, large values of $N_\mathrm{obs}$ appear in the top row referring to a Sun-like host star, intermediate values for the K star in the center row, and small values for the M dwarf host star in the bottom line. This trend is explained by the ratio of the satellite radius to the stellar radius. The moon's transit is much deeper in the M dwarf light curve than in the light curve of the Sun-like star, hence, fewer transits are required to detect it. Second, going from short to wide stellar distances, $N_\mathrm{obs}$ decreases. This decline is caused by the decreasing orbital velocity of the planet-moon binary around the star. In wider stellar orbits, transits have a longer duration, and so the binary's passage of the stellar disk yields more data points. The binned data then has smaller error bars and allows for more reliable $\chi^2$ fitting. Third, $N_\mathrm{obs}$ converges to a minimum value in the widest orbits. For a Sun-like star, this minimum number of transits is about 120 at 0.9\,AU and beyond, it is 34 for the $0.7\,M_\odot$ star at 0.5\,AU and beyond, and roughly a dozen transits beyond 0.4\,AU around a $0.4\,M_\odot$ star. In these regimes, white noise is negligible and the recovery of the injected moons depends mostly on the ratio $R_\mathrm{s}/R_\star$. $N_\mathrm{obs}$ is then comparable to the principled threshold $N_\mathrm{OSE}$ imposed by the OSE nature (see Section~\ref{sub:emergence}).

Translated into the required duty cycle of a telescope, timescales increase towards wider orbits, simply because the circumstellar orbital periods get longer. As an example, 107 transits were required in my simulations to discover a Ganymede-like satellite around a Jupiter-like planet orbiting a $0.7\,M_\odot$ star at 0.1\,AU, corresponding to a monitoring over 4\,yr. Only 34 transits were required for the planet-moon system at 0.5\,AU around the same star, but this means an observation time of 14.4\,yr. The odds of finding a Ganymede-sized moon transiting a G star in the available \textit{Kepler} data are poor, with $t_\mathrm{obs}>8.5$\,yr in any stellar orbit, and $t_\mathrm{obs}>100$\,yr beyond about 0.8\,AU. K stars are more promising candidates with $t_\mathrm{obs}$ as small as 4\,yr at a distance of 0.1\,AU. M dwarfs, finally, show the best prospects for Ganymede-like exomoons, because $t_\mathrm{obs}<4$\,yr at 0.2\,AU. Around the $0.4\,M_\odot$ star, this distance encases planet-moon binaries in the stellar HZ.

In the lower chart of Figure~\ref{fig:Nobs}, a planet with the 10-fold mass of Jupiter and an exomoon of $0.86\,R_\oplus$ is considered. In most cases, $N_\mathrm{obs}$ for a given star and stellar distance is smaller than in the left panel, because the moon is larger and causes a transit signal that is better distinguishable from the noise. But different from the left chart, $N_\mathrm{obs}$ does not strictly decrease towards low-mass stars. The $0.7\,M_\odot$ K star shows the best prospects for this exomoon's photometric OSE detection in the available \textit{Kepler} data. While a Sun-like host star could reveal the satellite after as few as 30 transits or 7.6\,yr at 0.4\,AU, the K star allows detection after 13 transits or 3.9\,yr at the same orbital distance, but still 27 transits or 10.7\,yr of observations would be required for the M star. The photometric OSE of such a hypothetical super-Ganymede moon could thus be measured in the available \textit{Kepler} data for planets as far as 0.4\,AU around K stars, thereby encompassing the stellar HZs.

Intriguingly, values of $N_\mathrm{obs}$ are larger in the bottom line of the lower panel than in the bottom line of the upper panel, at a given stellar distance. This is counterintuitive, as the larger satellite radius (lower chart) should decrease the number of required transits, in analogy to the Sun-like and K star cases. The discrepancy in the M dwarf scenario, however, is both an artifact of my boundary conditions, which require the fitted values of $R_\mathrm{s}$ and $a_{\mathrm{p}1}$ to deviate less than 10\,\% from the genuine values of the injected test moons, and the very nature of the photometric OSE. For very small stellar radii, such as the M dwarf host star, and relatively large satellite radii, such as the one used in the right chart, the OSE scatter becomes comparatively larger than observational noise effects. Hence, the K star in the center line of the lower panel of Figure~\ref{fig:Nobs} actually yields the most promising odds for the detection of super-Ganymede-like exomoons.

My choice of a 10\,\% deviation in both $R_\mathrm{s}$ and $a_\mathrm{ps}$ between the genuine injected exomoon and the best fit is arbitrary. To test its credibility, I ran a suite of randomized planet-only transits and corresponding $\chi^2$ fits for a Jupiter-like planet orbiting the G, K, and M dwarf stars at 0.5\,AU, respectively, which is roughly in the center of the top panel of Figure~\ref{fig:Nobs}. I generated 100 white-noisy phase-folded light curves after 151 (for the G star), 34 (K dwarf), and 13 (M dwarf) transits, corresponding to the number of transits required to gather 68\,\% of the genuine satellite systems within 10\,\% of the injected moon parameters in my previous simulations. After these additional $3\times100$ ``no moon'' runs, the corresponding best-fit distributions turned out to be almost randomly distributed in the $R_\mathrm{s}$-$a_\mathrm{ps}$ plane with a slight clustering toward smaller satellite radii, and no $10\,\%\times10\,\%$ bin contained more than a few of the best fits. In contrast, if a moon were present, the best-fit systems would be distributed according to a Gaussian distribution around the genuine radius and planetary distance of the moon. I conclude that a by-chance clustering within $10\,\%$ of any given location in the $R_\mathrm{s}$-$a_\mathrm{ps}$ plane is $\lesssim10^{-2}$. In turn, finding at least $68\,\%$ of the measurements within any $10\,\%\times10\,\%$ bin in the $R_\mathrm{s}$-$a_\mathrm{ps}$ plane makes a genuine moon system a highly probable explanation.

\section{Discussion}

\subsection{Methodological Comparison with other Exomoon Detection Techniques}

\subsubsection{TTV/TDV-based Exomoon Searches}

While TTV and TDV refer to variations in the planet's transit light curve, the photometric OSE directly measures the decrease in stellar brightness caused by one or multiple moons. Combined TTV and TDV measurements allow computations of the planet-satellite orbital semi-major axis $a_\mathrm{ps}$ and a moon's mass $M_\mathrm{s}$ \citep{2009MNRAS.396.1797K,2009MNRAS.392..181K}. All descriptions of the TTV/TDV-based search for exomoons are restricted to one-satellite systems. The photometric OSE enables measurements of $a_\mathrm{ps}$ and the satellite radii $R_\mathrm{s}$ in multiple exomoon systems. In comparison to the TTV/TDV method, the OSE can be measured with analytical expressions (Equation~(\ref{eq:OSE_multi}) for the photometric OSE; Equation~(\ref{eq:ProbDens_TDV}) for TDV-OSE; Equation~(\ref{eq:ProbDens_TTV}) for TTV-OSE). Note that TTV and TDV still need to be removed prior to analyses of the photometric OSE.

A major distinction between TTV and TDV correction for the purpose of OSE analysis, compared to the actual detection of an exomoon via its TTV and TDV imposed on the planet, lies in the irrelevance of the TTV/TDV origin for the photometric OSE analysis. On the contrary, for TTV/TDV-based exomoon searches these corrections need to be accounted for in a consistent star-planet-satellite model of the system's orbital dynamics to exclude perturbing planets as the TTV/TDV source. If such a dynamical model is not applied, then still $a_\mathrm{ps}$ and $M_\mathrm{s}$ can be inferred if both TTV and TDV can be measured \citep{2009MNRAS.392..181K,2009MNRAS.396.1797K} and a single moon is \textit{assumed} as the originator of the signal. Exomoon detection via photometric OSE, on the other hand, can be achieved with much less computational power, practically within minutes, as the transit light curve is phase-folded without TTV/TDV corrections. Determination of $a_\mathrm{ps}$ should be mostly unaffected, because TTV and TDV are supposed to be of the order of minutes \citep{2009MNRAS.400..398K,2013AsBio..13...18H,2013MNRAS.432.2549A}, whereas the photometric OSE extends as far as a few to tens of hours around the planetary transit (see Figures~\ref{fig:Keplerbin}Ê\, and \ref{fig:Keplerbinfit}), depending on the mass of the host star and the semi-major axis of the planet-satellite barycenter around the star. The satellite radius is even more robust against uncorrected TTV/TDV, since it is derived from the shape and depth of the OSE signal, not from its duration (see Figure~\ref{fig:zoom}).

Ultimately, the photometric OSE allows for the detection and characterization of multi-satellite systems, whereas currently available models ofÊ\, the TTV/TDV strategy cannot unambiguously disentangle the underlying satellite architecture of multiple satellite systems. Since the number of satellites around a giant planet is supposed to vary with planetary mass and depending on the formation scenario \citep{2010ApJ...714.1052S}, the photometric OSE is a promising alternative to TTV/TDV-based exomoon searches when it comes to understanding the formation history of extrasolar planets and moons.

In Sections~\ref{subsec:TDV-OSE} and \ref{subsec:TTV-OSE}, I examine the distribution of exomoon-induced TDV and TTV measurements. Combined with radial stellar velocity measurements and with the photometric OSE, they allow for a full parameterization of a star-planet-moon system. TTV-OSEs or TDV-OSEs alone may not unambiguously yield exomoon detections because they can be mimicked by planet-planet interactions \citep{2013ApJS..208...16M}. But if photometric OSEs indicate a satellite system, then TTV-OSE and TDV-OSE can be used to further strengthen the validity of the detection. In particular, TDV-OSE and TTV-OSE both offer the possibility of measuring a satellite's mass, which is unaccessible via the photometric OSE. In the spirit of Occam's razor, simultaneous observations of the photometric OSE, TTV-OSE, and TDV-OSE in longterm observations of a system would make an exomoon system the most plausible explanation, rather than a tilted transiting ring planet suffering planet-planet perturbations.

\subsubsection{Direct Photometric Exomoon Transits}

First, in comparison to direct observations of individual transits, the photometric OSE technique does not require dynamic modeling of the orbital movements of the star-planet-moon system, which drastically reduces the demands for computational power compared to photodynamic modeling \citep{2011MNRAS.416..689K}. Second, the amplitude of the photometricÊOSE signal in the phase-folded light curve is similar to the transit depth of a single exomoon transit, namely about $(R_\mathrm{s}/R_\star)^2$. But in contrast to single-transit analyses \citep[a method not applied by][by the way]{2011MNRAS.416..689K}, OSE comes with a substantial increase in signal-to-noise by averaging over numerous transits (see Figures~\ref{fig:Keplerobs} and \ref{fig:Keplerbin}). Third, OSE has a more complex imprint in the phase-folded light curve -- the ingress and egress patterns of the probability functions as well as a contribution to the total depth of the major transit trough (see Figure~\ref{fig:zoom}) -- and thereby offers a larger ``leverage'' to tackle moon detections more securely. However, speed is only gained in exchange for loss of information, which is reasonable for the most likely cases considered in this paper (coplanarity, circularity, masses separated by orders of magnitudes, radii and distances differing by at least an order of magnitude).

\subsubsection{Other Techniques}

Besides TTV/TDV measurements and direct photometric transit observations, a range of other techniques to search for exomoons have been proposed (see Section~\ref{sec:context}). In comparison to direct imaging of a planet-moon binary's photocentric wobble, which requires an angular resolution of the order of microarcseconds for a planet-moon binary similar to Saturn and Titan \citep{2007A&A...464.1133C}\footnote{Note that the authors use a mass ratio of 0.01 between Titan and Saturn to yield this threshold. However, the true mass ratio is actually about $2\times10^{-4}$, so the value for the required angular resolution might even be much smaller.}, the technical demands for photometric OSE measurements are much less restrictive. In other words, they are already available with the \textit{Kepler} telescope and the upcoming \textit{Plato 2.0} mission. Detections of planet-moon mutual eclipses require modeling of the system's orbital dynamics, which costs substantially more computing time than fitting Equation~(\ref{eq:OSE_multi}) to a phase-folded transit curve or Equations~(\ref{eq:ProbDens_TDV}) and (\ref{eq:ProbDens_TTV}) to the distribution of TDV and TTV measurements. Mutual eclipses can also be mimicked or blurred by star spots, and they are hardly detectable for moon's as small as Ganymede orbiting a Jovian planet. And referring to direct imaging of tidally heated exomoons, a giant planet with a spot similar to Jupiter's Giant Red Spot could also mimic a satellite eclipse.

Detections of the Rossiter-McLaughlin effect of a Ganymede-sized moon around a Jupiter-like planet requires accuracies in radial velocity measurements of the order of a few centimeters per second \citep{2010MNRAS.406.2038S} and will only be feasible for extremely quiet stars and with future technology \citep{2012ApJ...758..111Z}.

In comparison to detections with microlensing, observations of exomoon-induced OSEs are reproducible. Announcements of possible exomoon detections around free-floating giant planets in the Galactic Bulge show the obstacles of this technique and should be treated with particular skepticism. Not only are microlensing observations irreproducible, but also from a formation point of view, a moon with about half the mass of Earth cannot possibly form in the protosatellite disk around a roughly Jupiter-sized planet \citep{2006Natur.441..834C}.

Exomoons orbiting exoplanets around pulsars constitute a bizarre family of hypothetical moons, but as the first confirmed exoplanets actually orbit a pulsar \citep{1992Natur.355..145W}, they might exist. Analyses by \citet{2008ApJ...685L.153L} suggest that exomoons around pulsar planet PSR\,B1620$-$26\,b, if they exist, need to be at least as massive as about 5\,\% the planetary mass, leaving satellites akin to solar system moons undetectable. More generally, their time-of-arrival technique can hardly access moons as massive as Earth even in the most promising cases of planets and moons in wide orbits (Karen Lewis, private communication).

Direct imaging searches for extremely tidally heated exomoons also imply a yet unknown family of exomoons \citep{2013ApJ...769...98P}, where the satellite is at least as large as Earth and orbits a giant planet in an extremely close, eccentric orbit. Exomoon-induced modulations of a giant planet's radio emission require the moon (not the planet) to be as large as Uranus \citep{2013arXiv1308.4184N}, that is, quite big. Such a moon does also not exist in the solar system.

\subsubsection{Comparison of Detection Thresholds}

The detection limit of the combined TTV-TDV method using \textit{Kepler} data is estimated to be as small as $0.2\,M_\oplus$ for moons around Saturn-like planets transiting relatively bright M stars with \textit{Kepler} magnitudes $m_\textit{K} <11$ \citep{2009MNRAS.400..398K}. Note, however, that the TTV-TDV method is susceptible to the satellite-to-planet mass ratio $M_\mathrm{s}/M_\mathrm{p}$, not to the satellite mass in general. The HEK team achieves accuracies down to $M_\mathrm{s}/M_\mathrm{p}\approx5\,\%$ \citep{2013ApJ...770..101K,2013ApJ...777..134K,2014ApJ...784...28K,}. Using the correlation of exomoon-induced TTV and TDV on planets transiting less bright M dwarfs ($m_\textit{K}=12.5$) in the stellar HZs, \citet{2013MNRAS.432.2549A} find less promising thresholds of $8$ to $10\,M_\oplus$, and such an exomoon's host planet would need to be as light as $25\,M_\oplus$. Those systems would be considered planet binaries rather than planet-moon systems. \citet{2011EPJWC..1101009L} simulated TTVs caused by the direct photometric transit signature of moons and found that these variations could indicate the presence of moons as small as $0.75\,R_\oplus$, with this limit increasing towards wide orbital separations. A similar threshold has been determined by \citet{2012MNRAS.419..164S}, based on their scatter-peak method applied to \textit{Kepler} short cadence data.

Hence, depending on the actual analysis strategy of moon-induced TTV and TDV signals, and depending on the stellar apparent magnitude, planetary mass, and planet-moon orbital separation, a wide range of detection limits is possible. Most important, detection capabilities via TTV/TDV measurements as per \citet{2009MNRAS.400..398K} decrease for increasing planetary mass, which makes them most sensitive to very massive moons orbiting relatively light gas planets such as Saturn and Neptune. Yet, the most massive planets are predicted to host the most massive moons \citep{2006Natur.441..834C,2013AsBio..13..315W}.

In comparison, the photometric OSE presented in this paper is not susceptible to planetary mass and can detect Ganymede- or Titan-sized moons around even the most massive planets. This technique is thus well-suited for the detection of extrasolar moons akin to solar system satellites. What is more, the photometric OSE is the first method to enable the detection and classification of multi-satellite systems.

Rings of giant planets could mimic the OSE of exomoons. However, planets at distances ${\lesssim}1$\,AU will have small obliquities, or spin-orbit misalignments, due to the tidal interaction with the star. This ``tilt erosion'' \citep{2011A&A...528A..27H,2011OLEB...41..539H} will cause potential rings to be viewed edge-on during transits and so they will tend to be invisible. Also, a ring's transit signature will not generate an OSE, but its signal would look similar during every single transit. Analyses of the photometric scatter before the planetary ingress and after the planetary egress could be used to discriminate genuine exomoon systems from ring systems \citep{2012MNRAS.419..164S}. As a ring system does not induce TTVs or TDVs, measurements of the TTV-OSE and/or TDV-OSE can be used as an independent method to confirm exomoon detections.

My predictions of the number $N_\mathrm{obs}$ of transits required to discover the hypothetical one-satellite system around a giant planet may be overestimations, because I used a fixed data binning of 30\,min. This binning delivered the most reliable detections for the simulated Jupiter-satellite system in the HZ around a K star, but moon systems at different stellar orbital distances and around other host stars have different orbital velocities and their apparent trajectories differ in duration. Thus, more suitable data binning -- for example a 10\,min binning for short-period transiting planets -- will yield more reliable fittings than derived in this report. My simulations are thought to cover a broad parameter space rather than a best fit for each individual hypothetical star-planet-exomoon system.

\subsection{Red Noise}
While my noise model assumes white noise only (Section~\ref{sub:noise}), detrending real observations will have to deal with red noise \citep{2013MNRAS.430.1473L}. Instrumental effects such as CCD aging as well as red noise induced by stellar granulation and spots need to be removed or corrected for before the assumption of a light curve dominated by white noise becomes appropriate. In cases where red noise is comparable to white noise, $N_\mathrm{obs}$ and $t_\mathrm{obs}$ as presented in Figure~\ref{fig:Nobs} will increase substantially.

The results shown in Figure~\ref{fig:Nobs} do, however, still apply to a subset of photometrically quiet stars, such as the host star of transiting planet TrES-2b that has also been observed by \textit{Kepler} \citep{2011ApJ...733...36K}. \citet{2010ApJ...713L.155B} have shown that about every second K dwarf and about $16\,\%$ of the M dwarfs in the \textit{Kepler} sample are less active than the active Sun. \citet{2011ApJS..197....6G} found similar activity levels of K and M dwarfs but cautioned that the small sample of K and M dwarfs in the \textit{Kepler} data as well as contamination by giant stars could spoil these rates.

An OSE-based exomoon survey focusing on quiet stars will automatically tend to avoid spotted stars. If nevertheless present, clearly visible signatures of big star spots can be removed from the individual transits before the light curve is detrended and phase-folded. As long as these removals are randomized during individual transits, no artificial statistical signal will be induced into the phase-folded curve. But if the circumstellar orbital plane of the planet-satellite system were substantially inclined against the stellar equator and if the star had spot belts, then spot crossings would occur at distinct phases during each single transit \citep[see HAT-P-11 for an example,][]{2011ApJ...743...61S}. Such a geometry would strongly hamper exomoon detections via their photometric OSE.

\subsection{OSE detections with \textit{Plato 2.0}}

As the simulations in Section~\ref{sec:results} show, \textit{Kepler}'s photometry is sufficiently accurate to detect the  photometric OSE of transiting exomoons around relatively quiet K and M dwarfs with intrinsic stellar noise below about 20\,ppm. Hence, from a technological point of view, the \textit{Plato 2.0} telescope with a detector noise similar to that of \textit{Kepler} offers a near-future possibility to observe exoplanetary transits with similarly high accuracy.\footnote{In 2014 February, the \textit{Plato 2.0} mission has just been selected by ESA as its third medium-class mission within the Cosmic Vision program. Launch is expected around 2024.} However, even with arbitrarily precise photometry the number of observed transits determines the prospects of OSE detections. Given that \textit{Plato 2.0} is planned to observe two star fields for two to three years \citep{2013EPSC....8..707R}, Figure~\ref{fig:Nobs} suggests that this mission could just deliver as many transits as are required to enable photometric OSE detections around M and K stars. Yet, the results presented in this paper strongly encourage longterm monitoring over at least five years of a given star sample to allow exomoons to imprint their OSEs into the transit light curves. If the survey strategy of \textit{Plato 2.0} can be adjusted to observe one field for about five years or more, then the search for exomoons could become an additional science objective of this mission.

The \textit{Transiting Exoplanet Survey Satellite} (TESS), however, is planned to have an observing duty cycle of only two years, and it will observe a given star for 72\,days at most. Hence, TESS cannot possibly discover the photometric OSE of exomoons.

\section{Conclusions}
\label{subsec:conclusions}

This paper describes a new method for the detection of extrasolar moons, which I refer to as the Orbital Sampling Effect (OSE). It is the first technique that allows for reproducible detections of extrasolar multiple satellite systems akin to those seen in the solar system. The OSE appears in three flavors: (1) the photometric OSE (Section~\ref{sub:photoOSE}), (2) the TDV-OSE (Section~\ref{subsec:TDV-OSE}), and (3) the TTV-OSE (Section~\ref{subsec:TTV-OSE}). The photometric OSE can reveal the satellite radii in units of stellar radii as well as the planet-moon orbital semi-major axes, but it cannot constrain the satellite masses. TDV-OSE and TTV-OSE can both constrain the satellite mass. PhotometricÊ\, OSE, TDV-OSE, and TTV-OSE offer important advantages over other established techniques for exomoon searches because (1) they do not require modeling of the moons' orbital movements around the planet-moon barycenter during the transit, (2) planet-moon semi-major axes, satellite radii, and satellite masses can be measured or fit with analytical expressions (Equations~(\ref{eq:OSE_multi}), (\ref{eq:ProbDens_TDV}), (\ref{eq:ProbDens_TTV})), and (3) the photometric OSE is applicable to multi-satellite systems. TDV-OSE and TTV-OSE can also reveal the masses of moons in multi-satellite systems, but this parameterization is beyond the scope of this paper.

My simulations of photometric OSE detections with \textit{Kepler}-class photometry show that Ganymede-sized exomoons orbiting Sun-like stars cannot possibly be discovered in the available \textit{Kepler} data. However, they could be found around planets as far as 0.1\,AU from a $0.7$ solar mass K star or as far as 0.2\,AU from a $0.4\,M_\odot$ M dwarf. The latter case includes planet-moon binaries in the stellar HZ. Exomoons with the 10-fold mass of Ganymede and Ganymede-like composition (implying radii around $0.86\,R_\oplus$) are detectable in the \textit{Kepler} data around planets orbiting as far as 0.2\,AU from a Sun-like host star, 0.4\,AU from the K dwarf star, or about 0.2\,AU from the M dwarf. The latter two cases both comprise the respective stellar HZ. What is more, such large moons are predicted to form locally around super-Jovian host planets \citep{2006Natur.441..834C,2010ApJ...714.1052S} and are therefore promising targets to search for.

To model realistic light curves or to fit real observations with a photometric OSE model, stellar limb darkening needs to be included into the simulations (Heller et al., in preparation). Effects on $N_\mathrm{obs}$ are presumably small for planet-moon systems with low impact parameters, because the stellar brightness increases to roughly $60\,\%$ when the incoming moon has traversed only the first $5\,\%$ of the stellar radius during a transit \citep{2004A&A...428.1001C}. What is more, effects of red noise have not been treated in this paper, and so the numbers presented in Figure~\ref{fig:Nobs} are restricted to systems where either (1) the host star is photometrically quiet at least on a $\approx10$ hr timescale or (2) removal of red noise can be managed thoroughly. The prescriptions of the three OSE flavors delivered in this paper can be enhanced to yield the sky-projected angle between the orbital planes of the satellites and the diameter of the star. In principle, the photometric OSE allows measuring the inclinations of each satellite orbit separately. The effect of mutual moon eclipses will be small in most cases but offers further room for improvement. Ultimately, when proceeding to real observations, a Bayesian framework will be required for the statistical assessments of moon detections. As part of frequentists statistics, the $\chi^2$ method applied in this paper is only appropriate because I do not choose between different models since the injected moon architecture is known a priori.

Another application of the OSE technique, which is beyond the scope of this paper, lies in the parameterization of transiting binary systems. If not only the secondary constituent (in this paper the moon) shows an OSE but also the primary (in this paper the planet), then both orbital semi-major axes ($a_1$ and $a_2$) around the common center of mass can be determined. If the total binary mass $M_\mathrm{b}=M_1+M_2$ were known from stellar radial velocity measurements, then it is principally possible to calculate the individual masses via $a_1/a_2=M_2/M_1$ and substituting, for example, $M_1=M_\mathrm{b}-M_2$. This procedure, however, would be more complicated than in the model presented in this paper, because the center of the primary transit could not be used as a reference anymore. Instead, as both the primary and the secondary orbit their common center of mass, this barycenter would need to be determined in each individual light curve and used as a reference for phase-folding.

To sum up, the photometric OSE, the TDV-OSE, and the TTV-OSE constitute the first techniques capable of detecting extrasolar multiple satellite systems akin to those around the solar system planets, in terms of masses, radii, and orbital distances from the planet, with currently available technology. Their photometric OSE signals should even be measurable in the available data, namely, that of the \textit{Kepler} telescope. After the recent failure of the \textit{Kepler} telescope, the upcoming \textit{Plato 2.0} mission is a promising survey to yield further data for exomoon detections via OSE. To increase the likelihood of such detections, it will be useful to monitor a given field of view as long as possible, that is, for several years, rather than to visit multiple fields for shorter periods.

\appendix

\section{A. Parameterization of the planetary ingress}
\label{sec:ingress}

\begin{figure}[t]
  \centering
  \scalebox{0.9}{\includegraphics[angle=90]{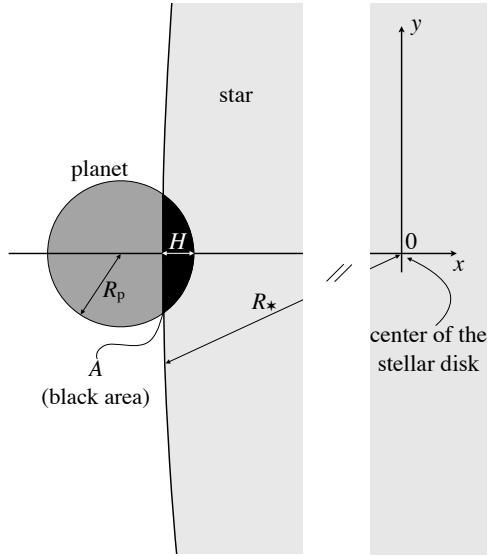}}\\
  \vspace{0.32cm}
  \caption{Parameterization of the planetary ingress.}
  \label{fig:ingress}
\end{figure}

During the ingress of the planet in front of the stellar disk, the planet blocks an increasing area $A$ of the stellar disk (black area in Fig.~\ref{fig:ingress}). With the star being substantially larger than the planet, the curvature of the stellar disk can be neglected and $A$ is determined by the height $H$ of the circular segment by

\begin{equation}
A = R_\mathrm{p}^2 \arccos\left(  1 - \frac{H}{R_\mathrm{p}} \right) - \wu{2R_\mathrm{p}H - H^2} (R_\mathrm{p}-H) \ \ .
\end{equation}

\noindent
In my simulations, $H=H(t)$ is a function of time. The temporary increase (during ingress) and decrease (during egress) of $A$ is visualized in Figure~\ref{fig:transit} by the gradual decrease in stellar brightness between about $-4$ and $-3$ hr and its gradual increase between $+3$ and $+4$ hr, respectively.

\section{B. Binning of simulated \textit{Kepler} data}
\label{sec:binning}

The set of simulated, noisy brightness measurements shown in Figure~\ref{fig:Keplerobs} is given by data points $b_i$. I divide the simulated observations into time intervals with running index $j$. The mean value of brightness measurements in the $j$th time bin is

\begin{equation}
\bar{b}_j=\frac{1}{K_j} \sum_{k=1}^{K_j} b_k \ \ ,
\end{equation}

\noindent
with $K_j$ as the number of data points $b_k$ in bin $j$. The variance $s_j^2$ within each bin is given by

\begin{equation}
s_j^2 = \frac{1}{K_j-1} \sum_{k=1}^{K_j}(b_k-\bar{b}_j)^2
\end{equation}

\noindent
and the standard deviation of the mean in that bin equals

\begin{equation}
\sigma_j = \frac{s}{\wu{K_j}} = \wu{ \frac{1}{K_j(K_j-1)} \sum_{k=1}^{K_j} (b_k-\bar{b}_j)^2 } \ \ .
\end{equation}

By increasing the bin width, say from 30 to 60\,min, it is possible to collect more data points per interval and to increase $K_j$, which in turn decreases $\sigma_j$ and increases accuracy. However, this comes with a loss in time resolution. The best compromise I found, at least for a satellite system on orbits similar to those of the Galilean moons but transiting in the HZ around a K star, is a 30\,min binning of long cadence \textit{Kepler} data. Transits of systems on wider circumstellar orbits have a longer duration, and thus might yield best results with a binning longer than 30\,min. Yet, their transits are less frequent, so this argument is only adequate for a comparable number of transits.

\section{C. $\chi^2$ minimization}
\label{sec:fitting}

I fit my simulated \textit{Kepler} observations of a one-satellite system with a brute force $\chi^2$ minimization technique, that is, I compute

\begin{equation}\label{eq:chisq}
\chi_{R_\mathrm{1},a_{\mathrm{p}1}}^2 = \frac{1}{K} \sum_{j=1}^{K} \frac{(b_j - m_j)^2}{\sigma_j^2}
\end{equation}

\noindent
for the whole parameter space and search for the global minimum. The parameter grid I explore spans $0.1\,R_\oplus~\leq~R_1~\leq~2\,R_\oplus$ in increments of $0.02\,R_\oplus$ and $2\,R_\mathrm{p}~\leq~a_{\mathrm{p}1}~\leq~30\,R_\mathrm{p}$ in steps of $0.1\,R_\mathrm{p}$. In Equation~(\ref{eq:chisq}), $K~=~\sum_{j}j$ is the number of binned data points to fit, $b_j$ denotes the binned simulated data points, and $m_j$ refers to the normalized brightness in bin $j$ predicted by the analytical model (Equation~(\ref{eq:OSE_multi})) for the satellite's radius $R_\mathrm{1}$ and semi-major axis $a_{\mathrm{p}1}$ to be tested (see Figure~\ref{fig:Keplerbinfit}).

\acknowledgments
I thank an anonymous reviewer for her or his valuable report. Karen Lewis' feedback also helped clarify several passages in this paper, and I thank Brian Jackson for his thoughtful comments. This work made use of NASA's ADS Bibliographic Services. Computations have been performed with {\tt ipython 0.13.2} on {\tt python 2.7.2} \citep{PER-GRA:2007}, and most figures were prepared with {\tt gnuplot 4.4} (\href{http://www.gnuplot.info}{www.gnuplot.info}).\\

\bibliography{ms}
\bibliographystyle{apj2}

\end{document}